\documentclass[10pt,a4paper]{article}
\usepackage{hyperref}
\usepackage{framed}

\usepackage{amsmath}
\usepackage{amssymb}

\begin{document}
\title{The role of Pryce's spin and coordinate operators in the theory of massive Dirac fermions}

\author{Ion I. Cot\u aescu \footnote{e-mail: i.cotaescu@e-uvt.ro}\\{\it West University of Timi\c soara, V. Parvan Ave. 4,}\\
	{\it Timi\c soara, 300223, Romania }}

\maketitle
\begin{abstract}
It is shown that the components of Pryce's spin operator of Dirac's theory are  $SU(2)$ generators of a representation carried by the space of  Pauli's spinors determining the polarization of the plane wave solutions of  Dirac's equation.  These operators are conserved via Noether theorem such that new conserved  polarization operators can be defined for various polarizations.  The corresponding one-particle operators of quantum theory are derived showing how these are related to the isometry generators of  the massive Dirac fermions of any polarization, including momentum-dependent ones. In this manner, the problem of separating conserved spin and orbital angular momentum operators is solved naturally. Moreover, the operator proposed by Pryce as mass-center coordinate is studied showing that after quantization this becomes in fact the dipole one-particle operator. As an example, the quantities determining the principal one-particle operators are derived for the first time in  momentum-helicity basis.

\end{abstract}

\section{Introduction}

The historical problem of finding a good spin operator of Dirac's theory comes from the fact that the Pauli spin part of the total angular momentum is not conserved separately  via Noether theorem. This problem was studied by many authors, giving rise to a rich literature, but without arriving to a commonly accepted solution (see for instance Refs. \cite{A,A1} and the literature indicated therein).  Nevertheless, there exist a privileged conserved self-adjoint spin operator satisfying all our exigences. This was found for the first time by Pryce \cite{A} in association with a mass-center  operator and re-defined later with the help of a Foldy-Wouthuysen transformation in momentum representation (${\bf p}$-rep.) \cite{FW} (presented in the Appendix A).

 Remaining outside this debate we tried to study how the polarization of Dirac's field can be changed by applying  suitable operators in configuration rep. (${\bf x}$-rep.). The polarization is determined by Pauli spinors that enter in the structure of the plane wave solutions of Dirac's equation,  offering supplemental  $SU(2)$ degrees of freedom but which are less studied so far. This is because of the difficulties in finding suitable operators able to transform the Pauli spinors only without affecting other quantities. Fortunately, we have found a spectral rep. of a class of integral operators allowing us to define such transformations (transfs.) acting in ${\bf x}$-rep. but changing the  Pauli spinors of the plane wave solutions in ${\bf p}$-rep..  The generators of these transfs.  close an $SU(2)$ algebra and, in addition, are conserved via Noether theorem just as the components (comps.) of the spin operator one looks for. As the action of these operators can be calculated in  ${\bf p}$-rep. we arrived at a surprising result: the  $SU(2)$ generators acting on Pauli's spinors  are just the comps. of the spin operator proposed by Pryce and re-defined then by Foldy and Wouthuysen in this rep..  In fact we found an alternative definition of the same spin operator but in a new framework allowing us to study how the principal operators of relativistic quantum mechanics (RQM) depend on polarization before and after quantization when these become the one-particle operators of the quantum field theory (QFT).   
 
 Our principal objective here  is to present the theory of  Dirac's free field in this framework  focusing on the role of polarization in determining the form,  action and physical meaning of principal operators of RQM and QFT. When the Pauli spinors depend on momentum,  as in the case of the largely used momentum-helicity basis,  we say that the polarization is {\em peculiar}. Otherwise we have  a {\em common} polarization, independent on momentum as, for example,  in the momentum-spin basis.  We present here general results concerning the principal operators for any polarization, first in RQM and then, after quantization, in QFT.   
 
The first novelty here is the spectral rep. allowing us to define  the spin operator  in ${\bf x}$-rep. as an integral operator having a kernel whose Fourier transform is just the Pryce spin operator in ${\bf p}$-rep.. Moreover, we give a new general definition of the polarization operator that holds even in the case of peculiar polarization getting a natural physical meaning  after quantization.   Similarly, we derive the action in ${\bf x}$-rep. of the associated  coordinate operator defined by Pryce in ${\bf p}$-rep. pointing out that this depend linearly on time and defining the corresponding velocity operator. Furthermore, performing the quantization we obtain for the first time the one-particle spin and polarization operators  as well as the isometry generators for any peculiar polarization. We show that these operators   depend on new Pauli-type momentum-dependent matrices and  'covariant' momentum derivatives. Moreover, we point out that the total angular momentum operator of QFT is split into a new orbital angular momentum and spin operators, each one being conserved separately.
However, the surprise is the operator defined by Pryce as mass-center position vector which becomes after quantization a time-dependent {\em dipole} operator whose velocity is known as the classical current.  The final original results we present here are the aforementioned matrices and derivatives determining the form of the one-particle operators  in momentum-helicity basis. 
  
We start in the next section revisiting briefly the covariant Dirac free field,  its symmetries  and the relativistic scalar product in ${\bf x}$-rep.. In Sec. 3,  devoted to ${\bf p}$-rep., we present first the general properties of the mode spinors of this rep. focusing then on the equivalence of  the covariant rep. with an orthogonal sum of a pair of Wigner's unitary and irreducible ones, induced by the rep. of spin half of  the $SU(2)$ group.   Here we obtain the structure of the mode spinors of any peculiar polarization and the transfs. of the wave functions in ${\bf p}$-rep. showing how  the generators of covariant and induced reps. are related among themselves. In the first part of Sec. 4 we introduce our new spectral rep.   helping us to define the spin and polarization operators in the next part and to analyze the associated coordinate operator in the last one.  

These results are obtained  in ${\bf x}$ and ${\bf p}$ reps. of RQM where there are problems in interpreting the anti-particle terms. By  a lucky chance, the framework adopted here allows us to apply easily the Bogolyubov method of quantization transforming the expectation values of  the operators of RQM into the one-particle operators of QFT. In Sec. 5, devoted to this procedure,  we give the principal one-particle operators showing how the  total angular momentum  is split into spin and orbital angular momentum conserved one-particle operators. The last part of this section is devoted to the associated coordinate operator that after quantization becomes the dipole one-particle operator evolving linearly in time thanks to the conserved operator of classical current. 

 In  Sec. 6 we give as example the operators of QFT in  momentum-helicity basis for which we derive for the first time the aforementioned Pauli-type matrices and momentum derivatives. Moreover, we discuss the difference between our polarization operator and the helicity one showing that these operators give eigenvalues with opposite signs in the antiparticle sector.   Finally we show how look the isometry generators when we turn back to RQM but constructed as the one-particle restriction of QFT. We observe that in the space of Pauli spinors defining one-particle states in  ${\bf p}$-rep. the spin and polarization operators of the relativistic approach are just the original Pauli ones of the non-relativistic theory.  
 
 The last section in which we present our concluding remarks is followed by two Appendices. In the first one we give some technical details concerning  the boost matrices,  projection operators and the Foldy-Wouthuysen transf. laying out the Pryce spin operator. In the second Appendix we discuss briefly the role of  the induced reps. in RQM.

\section{Massive Dirac field}

Let us start with the Minkowski space-time, $(M,\eta)$, having the metric $\eta={\rm diag}(1,-1,-1,-1)$ and Cartesian coordinates $x^{\mu}$ ($\alpha,\,\beta,...\mu,\,\nu...=0,1,2,3$). The isometries of $M$, are transfs. of the  Poincar\' e group  ${\cal P}_{+}^{\uparrow} =T(4)\,\circledS\, L_{+}^{\uparrow}$ \cite{WKT}, $(\Lambda,a): x\to x'=\Lambda x+a$, formed by transfs. $\Lambda \in L_{+}^{\uparrow}$ of the  orthochronous proper Lorentz group,  preserving the metric $\eta$, and four dimensional translations $a\in T(4)$. The universal covering group of the Poincar\' e one,  $ \tilde{\cal P}^{\uparrow}_{+}=T(4)\,\circledS\, SL(2,\mathbb{C})$, includes transfs.  $\lambda\in SL(2,\mathbb{C}) $ related to  those of the Lorentz group  through the canonical homomorphism, $\lambda\to \Lambda(\lambda)$ \cite{WKT}. 

The {\em covariant} Dirac field, $\psi  :\, M\to {\cal V}_D  $, is locally defined over $M$ with values in the vector spaces ${\cal V}_D  $ carrying  the finite-dimensional  rep. $\rho_D=(\frac{1}{2},0)\oplus(0,\frac{1}{2})$ of the  $SL(2,\Bbb C)$ group where one defines the Dirac  $\gamma$-matrices that  satisfy  $\{\gamma^{\mu},\gamma^{\nu}\}=2\eta^{\mu\nu}$  giving rice to $SL(2,\mathbb{C})$ generators.   Here we consider exclusively  the chiral representation (with diagonal $\gamma^5$) in which the transfs.   
\begin{equation}
	\lambda(\omega)=\exp\left(-\frac{i}{2}\omega^{\alpha\beta}s_{\alpha\beta}\right)\in \rho_D\,, \quad
		s^{\mu\nu}=\frac{i}{4}\left[\gamma^{\mu},\gamma^{\nu}\right]\,,  
\end{equation}
with  real-valued parameters, $\omega^{\alpha \beta}=-\omega^{\beta\alpha}$,  are reducible to the  subspaces of the irreducible reps. $(\frac{1}{2},0)$ and $(0,\frac{1}{2})$ of $\rho_D$ \cite{WKT,Th}. We denote by $r={\rm diag}(\hat r,\hat r)\in\rho_D\left[SU(2)\right]$  the transfs. we call here simply rotations, for which we use the Cayley-Klein parameters $\theta^i=\frac{1}{2}\epsilon_{ijk}\omega^{jk}$ and  the generators $s_i= \frac{1}{2}\epsilon_{ijk}s^{jk}  ={\rm diag}(\hat s_i,\hat s_i)$,  where $\hat s_i=\frac{1}{2}\sigma_i$ are the comps. of the Pauli spin operator depending on Pauli's matrices $\sigma _i$  ($i,\,j,\,k... = 1,2,3$).  Similarly,  the transfs.  $l={\rm diag}(\hat l,\hat l^{-1})\in \rho_D\left[ SL(2,\mathbb{C})/SU(2)\right]$,  called here the Lorentz boosts, are generated by the matrices $s_{0i}={\rm diag}(i\hat s_i, -i\hat s_i)$ whose  parameters are denoted by $\tau^i=\omega^{0i}$. Summarizing we may write
\begin{eqnarray}
r(\theta)&=&{\rm diag}(\hat r(\theta),\hat r(\theta))\,,\quad~~ \hat r(\theta)=e^{-i \theta^i \hat s_i}=e^{-\frac{i}{2} \theta^i \sigma_i}\,, \label{r}\\
l(\tau)&=&{\rm diag}(\hat l(\tau),\hat l^{-1}(\tau))\,,\quad \hat l(\tau)=e^{ \tau^i \hat s_i}=e^{\frac{1}{2} \tau^i \sigma_i} \,.\label{l}
\end{eqnarray}
In the covariant parametrization  the associated Lorentz transfs. may be expanded as 
\begin{equation}\label{Lala}
\Lambda^{\mu\,\cdot}_{\cdot\,\nu}[\lambda(\omega)]=\delta^{\mu}_{\nu}
+\omega^{\mu\,\cdot}_{\cdot\,\nu}+\omega^{\mu\,\cdot}_{\cdot\,\alpha}\omega^{\alpha,\cdot}_{\cdot\,\nu}+\cdots	\,,
\end{equation}
as it results from the canonical homomorphism \cite{WKT} .

The massive Dirac field $\psi$  of mass $m$ and its Dirac adjoint,  $\overline{\psi}=\psi^+\gamma^0$, are canonical variables of the action 
\begin{equation}\label{actionD}
	{\cal S}[\psi]=\int d^4 x {\cal L}_D(\psi, \overline{\psi})\,,
\end{equation}
defined by  the Lagrangian density, 
\begin{equation}
	{\cal L}_D(\psi)=\frac{i}{2}\,[\overline{\psi}\gamma^{\alpha}\partial_{\alpha}\psi-
	(\overline{\partial_{\alpha}\psi})\gamma^{\alpha}\psi] - 	m\overline{\psi}\psi\,.
\end{equation}
This action gives rise to the Dirac equation,
\begin{equation}\label{E}
E_D\psi=(i\gamma^{\mu}\partial_{\mu}-m)\psi=0	\,,
\end{equation}  
and the form of the relativistic scalar product 
\begin{equation}\label{sp}
	\langle\psi,\psi'\rangle_D=\int d^3x\overline{\psi}(x)\gamma^0\psi'(x)=\int d^3x \psi^ {+} (x)\psi'(x)	\,,
\end{equation}
related to the conserved  electric charge. 

The Dirac field transforms   under isometries according to the {\em covariant} rep.  $T   \,:\,(\lambda, a)\to  T_{\lambda,a}$ of the group $ \tilde{\cal P}^{\uparrow}_{+}$, as \cite{WKT}
\begin{equation}\label{TAa}
	(T_{\lambda,a}\psi  )(x)
	=\lambda \psi  \left(\Lambda(\lambda)^{-1}(x-a)\right)\,.
\end{equation}
The well-known basis-generators of  this rep., 
\begin{eqnarray}
	P_{\mu}=-\left.i\frac{\partial T_{1,a}}{\partial a^{\mu}}\right|_{a=0}\,, \quad 
	J_{\mu\nu}=\left.i\frac{\partial T_{\lambda(\omega),0}}{\partial \omega^{\mu\nu}}\right|_{\omega=0}\,,
\end{eqnarray}
 may be  rewritten in vector notation, separating the momentum comps. and energy  operator,  $ P^i=-i\partial_i$ and $ H=P_0=i\partial_t$, and denoting  the $SL(2,{\Bbb C})$ generators as, 
\begin{eqnarray}
	J_i  &=&\frac{1}{2}\,\varepsilon_{ijk}	J  _{jk}=
	-i\varepsilon_{ijk}\underline{x}^j\partial_k+s_i  \,,\label{J}\\
	K_i  &=& J_{0i}  =i (\underline{x}^i
	\partial_t+t\partial_i)+s_{0i}  \,, \label{K}
\end{eqnarray}
where the comps. $\underline{x}^i$ of  the coordinate vector-operator $\underline{\bf x}$  act as $(\underline{x}^i \psi)(x)=x^i\psi(x)$. The set  $\{ H,  P^i, J  _i, K  _i\}$ represents the usual basis of the Lie algebra ${\rm Lie}(T)  $ of  the rep. $T  $ \cite{WKT}. 

The action (\ref{actionD}) is  invariant under isometries such that  the scalar product (\ref{sp}) is also  invariant,  
\begin{equation}
\langle T  _{A,a}\psi  ,T  _{A,a}\psi'  \rangle_D=\langle\psi ,\psi' \rangle_D\,,	
\end{equation}
because  the generators $X\in {\rm Lie}(T)$  are self-adjoint,  obeying  $\langle X\psi  ,\psi'  \rangle_D=\langle\psi, X\psi'  \rangle_D$,  as the  $SL(2,{\Bbb C})$ generators of the rep. $\rho_D$ are Dirac self-adjoint, $\overline{s}_{\mu\nu}={s}_{\mu\nu}$. All these generators  are conserved via Noether's theorem in the sense that their expectation values $\langle \psi  , X\psi  \rangle_D$ are  independent on time. Therefore, we may conclude that in this framework the covariant rep. $T$ behaves as an {\em unitary} one with respect to the relativistic scalar product  (\ref{sp}).

Of a special interest is the total angular momentum operator ${\bf J}=\underline{\bf x}\land{\bf P} +{\bf s}$, defined by Eq. (\ref{J}), which  is formed by the orbital term $\underline{\bf x}\land{\bf P}$ associated to the reducible Pauli spin operator ${\bf s}$. Unfortunately, these operators are not conserved separately in the sense of  above definition such that they do not have a correct physical meaning in special relativity. For this reason, one seeks for a {\em conserved} spin operator ${\bf S}$ associated to a suitable coordinate operator, ${\bf X}$, allowing the new splitting
\begin{equation}\label{spli}
{\bf J}=\underline{\bf x}\land{\bf P} +{\bf s}={\bf X}\land {\bf P}+{\bf S}	\,,
\end{equation}
whose orbital term, ${\bf X}\land {\bf P}$, has to be also conserved.  This is the principal problem we discuss in this  paper focusing on Pryce's spin and coordinate operators.  

The invariants of the Dirac field are  the eigenvalues of  Casimir operators of the rep. $T$ that read \cite{Th}
\begin{eqnarray}
	{C}_1&=& P_{\mu} P^{\mu}\sim m^2\,,\label{CM1}\\
	{C}_2  &=&-\eta_{\mu\nu} W^{\mu}	W^{\nu} \sim m^2 s(s+1)\,,\quad s=\textstyle{ \frac{1}{2}}\,,\label{CM2}
\end{eqnarray}
where the Pauli-Lubanski operator \cite{WKT},
\begin{equation}\label{PaLu}
	W^{\mu}=-\frac{1}{2}\,\varepsilon^{\mu\nu\alpha\beta}P_{\nu}  J_{\alpha\beta}  \,,
\end{equation}
has the components
\begin{equation}
	W  _0={ J}  _i{ P}^i={s} _i{P}^i\,,\quad  W  _i= H\,{ J}_i  +
	\varepsilon_{ijk}{ P}  _j { K}_k\,,
\end{equation}
as we use  $\varepsilon^{0123}=-\varepsilon_{0123}=-1$.  These operators are  considered by many authors as the comps. of a covariant four-dimensional spin operator of RQM as long as $W_0$ is just the helicity operator \cite{C}. This  is the only differential operator able to define a  polarization but  leading to minor difficulties at the level of QFT as we shall see later. For this reason we believe that a more convenient polarization operator must come from the algebra ${\rm Lie}(T)$. 

\section{Momentum representation}

 The Dirac equation allows plane wave solutions that form a basis of ${\bf p}$- rep.  in which the commuting operators  $H$ and $P^i$  are diagonal. This system of operators is incomplete as a polarization operator is missing. This  will be added after discussing the problem of finding conserved spin and polarization operators. For the time being we assume that the arbitrary Pauli spinors entering  in the structure of  the rest frame plane waves define implicitly the polarization.
 
 \subsection{Frequecies separation}
 
 The general solutions of the free Dirac equation may be expanded in terms of  usual  mode spinors (or fundamental spinors),  $U_{{\bf p},\sigma}$ and  $V_{{\bf p},\sigma}=C  U_{{\bf p},\sigma}^*$, of positive and respectively negative frequencies,   related through the charge conjugation defined by the matrix $C=C^{-1}=i\gamma^2$  and acting as \cite{BDR}
 \begin{equation}
{\gamma^{\mu}}^*=-C\gamma^{\mu}C~\to~{s}_{\mu\nu}^*=-C s_{\mu\nu}C~\to~	{\lambda}^*=C \lambda C \,.	
 \end{equation} 
The plane wave mode spinors   satisfy the Dirac equation and the eigenvalues problems,  
\begin{eqnarray}
	& H U_{{\bf p},\sigma}=E(p) U_{{\bf p},\sigma}\,,\quad ~~
&	H V_{{\bf p},\sigma}=-E(p) V_{{\bf p},s\sigma}\,,\\
	&{ P}^i U_{{\bf p},\sigma}={p}^i\, U_{{\bf p},\sigma}\,,\quad ~~~~~
	&	{ P}^i V_{{\bf p},\sigma}=-{p}^i\, V_{{\bf p},\sigma}\,,
\end{eqnarray}
where $E(p)=\sqrt{m^2+{p}^2}$   ($p=|{\bf p}|$) is the relativistic energy. Then the general solution of the Dirac equation, we call here simply the Dirac freee field, can be expanded as \cite{KH,BDR}
\begin{eqnarray}\label{Psi}
	\psi  (x)=	\psi  ^+(x)+	\psi  ^-(x)=\int d^3p \sum_{\sigma}\left[U_{{\bf p},\sigma}(x) \alpha_{\sigma}({\bf p}) +V_{{\bf p},\sigma}(x) \beta^ {*} _{ \sigma}({\bf p})\right]\,,
\end{eqnarray}
in terms of mode functions in ${\bf p}$-rep.,  $\alpha_{\sigma}$ and $\beta_{\sigma}$,  of  particles and  respectively antiparticles  of arbitrary {\em polarization} $\sigma=\pm\frac{1}{2}$ that may be defined in various manners as we shall show  after studying the spin operators.  In this manner, the space of Dirac's mode spinors, ${\cal F}_D={\cal F}^+_D\oplus {\cal F}^-_D$,  is split in two orthogonal subspaces of mode spinors  of positive and respectively negative frequencies.

The mode spinors  prepared at the initial time $t_0=0$ have the general form
\begin{eqnarray}
	U_{{\bf p},\sigma}(x)&=&u_{\sigma}({\bf p})\frac{1}{(2\pi)^{\frac{3}{2}}} \,e^{-iE(p)t+i{\bf p}\cdot{\bf x}}\,,\label{U} \\
	V_{{\bf p},\sigma}(x)&=&v_{\sigma}({\bf p})\frac{1}{(2\pi)^{\frac{3}{2}}}\, e^{iE(p)t-i{\bf p}\cdot{\bf x}}\,,\label{V}
\end{eqnarray} 
where  the spinors $u_{\sigma}({\bf p})$ and $v_{\sigma}({\bf p})=C  u_{s\sigma}({\bf p})^*$ must be normalized in order to obtain the orthonormalization relations 
\begin{eqnarray}
	\langle U_{{\bf p},\sigma}, U_{{{\bf p}\,}',\sigma'}\rangle_D &=&
	\langle V_{{\bf p},\sigma}, V_{{{\bf p}\,}',\sigma'}\rangle_D =	\delta_{\sigma\sigma^{\prime}}\delta^{3}({\bf p}-{\bf p}\,^{\prime})\,,\label{ortU}\\
	\langle U_{{\bf p},\sigma}, V_{{{\bf p}\,}',\sigma'}\rangle_D &=&
	\langle V_{{\bf p},\sigma}, U_{{{\bf p}\,}',\sigma'}\rangle_D =0\,, \label{ortV}
\end{eqnarray}
and the completeness condition,
\begin{eqnarray}
\int d^3p \sum_{\sigma}\left[ U_{{\bf p},\sigma}(t,{\bf x})U_{{\bf p},\sigma}^{+}(t,{\bf x}')+V_{{\bf p},\sigma}(t,{\bf x})V_{{\bf p},\sigma}^{+}(t,{\bf x}')  \right]=\delta^3({\bf x}-{\bf x}')\,.\label{comp}
\end{eqnarray}
Moreover, Eqs. (\ref{ortU}) and (\ref{ortV}) help us to write the inversion formulas
\begin{equation}\label{inv}
	\alpha_{\sigma}({\bf p})=\langle U_{{\bf p},\sigma},\psi \rangle_D\,, \quad 
		\beta_{\sigma}({\bf p})=\langle \psi,  V_{{\bf p},\sigma} \rangle_D\,, 
\end{equation}
we need in applications.

In RQM the physical meaning of the field $\psi$ is encapsulated in the wave functions in ${\bf p}$-rep., $\alpha_{\sigma}$ and $\beta_{\sigma}$, that form the Pauli spinors, 
\begin{equation}\label{alpha}
\alpha=\left( 
\begin{array}{l}
\alpha_{\frac{1}{2}}\\
 \alpha_{-\frac{1}{2}} 
\end{array}\right)	\in {\cal F}_{\alpha}\,,
\quad
\beta=\left( 
\begin{array}{llc}
	\beta_{\frac{1}{2}}\\
	\beta_{-\frac{1}{2}} 
\end{array}\right) 	\in {\cal F}_{\beta}\,,
\end{equation}
of the Hilbert spaces ${\cal F}_{\alpha}\sim {\cal F}_{\beta}$ equipped with scalar products, 
\begin{eqnarray}\label{spa}
\langle \alpha, \alpha'\rangle=\int d^3p \,\alpha^+({\bf p})\alpha'({\bf p})=\int d^3p \sum_{\sigma}\alpha_{\sigma}^*({\bf p})	\alpha_{\sigma}'({\bf p})\,,
\end{eqnarray}
and similarly for the spinors $\beta$,  such that we can write
\begin{equation}\label{spp}
\langle \psi  , \psi  '\rangle_D =\langle \alpha,\alpha'\rangle + \langle \beta,\beta'\rangle	\,,
\end{equation}
after using  Eqs.  (\ref{ortU}) and (\ref{ortV}).

We remind the reader that the differential operators in ${\bf x}$-rep., $F(i\partial_{\mu}):  {\cal F}_D^{\pm} \to  {\cal F}_D^{\pm}$,   give rise to multiplicative operators  in ${\bf p}$-rep., $\tilde F(p^{\mu})$, acting differently on the mode spinors $U$ and $V$,   
\begin{eqnarray}
\left[	F(i\partial_{\mu}) \psi\right](x)=\int d^3p \sum_{\sigma}\left[ \tilde F({p^{\mu}})U_{{\bf p},{\sigma}}( x)\alpha_{\sigma}({\bf p}) +\tilde F(-{p^{\mu}})V_{{\bf p},{\sigma}}(x)\beta^*_{\sigma}({\bf p})\right]\,.\label{F2}
\end{eqnarray}
We say  that these operators are diagonal on ${\cal F}_D$ as they do not mix the mode spinors of different frequencies. For example, the Dirac Hamiltonian operator $H_D=-i\gamma^0\gamma^i\partial_i+m\gamma^0$ acts as 
\begin{eqnarray}
	(H_D U_{{\bf p},\sigma})(x)&=&\tilde{H}_D({\bf p})U_{{\bf p},\sigma}(x)=E(p) U_{{\bf p},\sigma}(x)\,,\label{HDU}\\
	(H_DV_{{\bf p},\sigma})(x)&=&\tilde{H}_D(-{\bf p})V_{{\bf p},\sigma}(x)=-E(p) V_{{\bf p},\sigma}(x)\,,\label{HDV}
\end{eqnarray}
where 
\begin{equation}\label{HDp}
\tilde H_D({\bf p})=m\gamma^0+\gamma^0\gamma^i p^i =E(p)\left[ \tilde	\Pi_+({\bf p})-\tilde	\Pi_-({\bf p})\right]	\,,
\end{equation}
is the Hamiltonian operator in ${\bf p}$-rep.  expressed in terms of the projection operators (\ref{Pip}) and (\ref{Pim}).

\subsection{Relating covariant and induced representations}

For developing our approach we need to work simultaneously in ${\bf x}$ and ${\bf p}$ reps. relating the covariant rep. of ${\bf x}$-rep. to the Wigner induced reps. transforming the spinors (\ref{alpha}). Moreover the Wigner method allows us to construct the mode spinors separating the Pauli spinors whose degrees of freedom have to be studied here.   

The wave functions of ${\bf p}$-rep. (\ref{alpha}) are defined on  orbits in momentum space, $\Omega_{\mathring p}=\{{\bf p}| {\bf p}=\Lambda \mathring{p}, \Lambda\in L_{+}^{\uparrow} \}$, that may be built by applying Lorentz transfs. on a {\em representative} momentum $\mathring{p}$ \cite{Wig, Mc}.  In the case of massive particles, we discuss here,  the {representative} momentum is just the rest frame one,  $\mathring{p}=(m,0,0,0)$.  The rotations that leave $\mathring{p}$ invariant, $R\mathring{p}=\mathring{p}$, form the {\em stable} group $SO(3)\subset L_{+}^{\uparrow}$ of $\mathring{p}$ whose  universal covering group $SU(2)$  is called the {\em little} group associated to the representative momentum $\mathring{p}$. 

For each momentum ${\bf p} $ there exist a set of  transfs. $L_{{\bf p}}$ generating it  as  ${\bf p}=L_{{\bf p}}\,\mathring{p}$. These transfs. are  defined up to arbitrary  rotations $R({\bf p})$ which  may depend on ${\bf p}$  as these do not change the representative momentum, $L_{{\bf p}}R({\bf p })\,\mathring{p}=L_{{\bf p}}\,\mathring{p}$.  This means that the orbit $\Omega_{\mathring p}$ is in fact an  homogeneous space $L_{+}^{\uparrow}/SO(3)$.  Consequently, the corresponding transfs.  $\lambda_{{\bf p}}\in \rho_D$  which satisfy $\Lambda(\lambda_{{\bf p}})=L_{{\bf p}}$ and $\lambda_{{\bf p}=0}=1\in \rho_D$ have the general form  $\lambda_{{\bf p}}= l_{\bf p} r({\bf p})$, being constituted by genuine Lorentz boosts $l_{\bf p}\in \rho_D[SL(2,{\Bbb C})/SU(2)]$ defined by Eq. (\ref{Ap})  and arbitrary rotations $r({\bf p})\in \rho_D[SU(2)]$ satisfying  $r({\bf p}=0)=1\in \rho_D$.  

Hereby we see that the functions (\ref{alpha}) are defined on the orbit $\Omega_{\mathring p}$ where the invariant measure is \cite{WKT}
\begin{equation}\label{measure}
\mu({\bf p})=\mu(\Lambda{\bf p})=\frac{d^3p}{E(p)}\,, \quad \forall \Lambda\in L_{+}^{\uparrow}\,. 	
\end{equation}
Therefore,   $\sqrt{E}\alpha$ and   $\sqrt{E}\beta$ are square integrable functions of the Hilbert space ${\cal L}^2(\Omega_{\mathring p}, \mu, {\cal V}_P)$ equipped with an invariant scalar product that can be put in the form (\ref{spa}). In practice it is convenient to work with the spaces ${\cal F}_{\alpha}\sim {\cal F}_{\beta}\sim {\cal L}^2(\Omega_{\mathring p}, d^3p, {\cal V}_P)$ which are isometric with ${\cal L}^2(\Omega_{\mathring p}, \mu, {\cal V}_P)$.

For investigating how these functions transform under  isometries one assumes that there exist a rep. $\tilde T :(\lambda,a)\to \tilde T_{\lambda,a}$, carried by the spaces  ${\cal F}_\alpha$ and ${\cal F}_\beta$, associated to the rep. $T$ as \cite{WKT,W,Th}
\begin{eqnarray}\label{basic}
(T_{\lambda,a}\psi)(x)=\int d^3p \sum_{\sigma}\left[U_{{\bf p},\sigma}(x)(\tilde T_{\lambda,a}\, \alpha)_{\sigma}({\bf p})+V_{{\bf p},\sigma}(x) (\tilde T_{\lambda,a}\,\beta)^ {*} _{ \sigma}({\bf p})\right]\,.
\end{eqnarray}
Taking into account that the covariant rep. $T$ is defined by Eq. (\ref{TAa}), we use the identity $(\Lambda x)\cdot p=x\cdot (\Lambda^{-1}p)$ and the invariant measure, $d^3pE(p)^{-1}=d^3p'E(p')^{-1}$, for changing the integration variable,
\begin{equation}\label{pLp}
{\bf p}~\to~	{{\bf p}\,}'=\Lambda(\lambda)^{-1}{\bf p} \,,	
\end{equation}
finding  the action of the operators $\tilde T$ \cite{WKT,Th}, 
\begin{eqnarray}
	&&\sum_{\sigma'}u_{\sigma'}({\bf p})(\tilde T_{\lambda,a}\,\alpha)_{ \sigma'}({\bf p})\frac{E(p')}{E(p)}\sum_{\sigma} \lambda u_{\sigma}\left( {{\bf  p}}' \right)\alpha_{\sigma}\left({{\bf p}}'\right) e^{i {a}\cdot{p}}\,,\\
	&&\sum_{\sigma'}v_{\sigma'}({\bf p})(\tilde T_{\lambda,a}\, \beta^ {*}) _{\sigma'}({\bf p})= \frac{E(p')}{E(p)}\sum_{\sigma} \lambda v_{\sigma}\left({{\bf p}}'\right)\beta^ {*} _{\sigma}\left( {\bf p}'\right) e^{-i {a}\cdot{p}}\,,
\end{eqnarray}
where $a\cdot p=a_{\mu}p^{\mu}=E(p)a^0-{\bf p}\cdot {\bf a}$.  

Furthermore, according to Wigner's general method, we introduce the spinors \cite{Th}, 
\begin{eqnarray}
	u_{\sigma}({\bf p})&=&N({p})\lambda_{{\bf p}}\,  \mathring{u}_{\sigma}=N(p)l_{{\bf p}}r({\bf p})\,  \mathring{u}_{\sigma}\,,  \label{WigU} \\  
		v_{\sigma}({\bf p})&=&C  u_{\sigma}^*({\bf p})=N(p)\lambda_{\bf p} \mathring{v}_{\sigma}=N(p)l_{\bf p}r({\bf p}) \mathring{v}_{\sigma}\,,   \label{WigV}
\end{eqnarray}
where  $N(p)\in\mathbb{R}	$  satisfying $N(0)=1$ is a normalization factor.  The rest frame spinors  $\mathring{u}_{\sigma}=u_{\sigma}(0)$ and $\mathring{v}_{\sigma} =v_{\sigma}(0)=C\mathring{u}_{\sigma}^*$  are solutions of the Dirac equation in the rest frame where this equation reduces to the eigenvalues problems of  the matrix $\gamma^0$,
\begin{equation}\label{cucu1}
	\gamma^0\mathring{u}_{\sigma}=\mathring{u}_{\sigma}\,, \quad  \gamma^0\mathring{v}_{\sigma}=-\mathring{v}_{\sigma}. 
\end{equation}
Then  the Wigner spinors (\ref{WigU}) and (\ref{WigV}) are solutions of the Dirac equation in any frame of ${\bf p}$-rep.. Indeed,   observing that the matrix $\gamma p=E(p)\gamma^0-{\gamma}^i p^i$ satisfies $\gamma p\,\lambda_{{\bf p}}=m \lambda_{{\bf p}}\gamma^0$  we obtain the Dirac equations in ${\bf p}$-rep.,
\begin{equation}\label{mucu1}
	(\gamma p-m)u_{\sigma}({\bf p})=0\,, \quad (\gamma p+m)v_{\sigma}({\bf p})=0\,,
\end{equation}
after exploiting Eqs. (\ref{cucu1}).

The matrices $\frac{1\pm \gamma^0}{2}$ form an orthogonal system of projection matrices such that the spinor subspaces  $\frac{1+ \gamma^0}{2}{\cal V}_D$ and $\frac{1- \gamma^0}{2}{\cal V}_D$ are orthogonal. Moreover,  we assume that all these spinors  are normalized, ${\mathring{u}}^+_{\sigma}\mathring{u}_{\sigma'}={\mathring{v}}^+_{\sigma}\mathring{v}_{\sigma'}=\delta_{\sigma\sigma'}$,  forming  complete systems on their subspaces,
\begin{equation}\label{ortuv}
	\sum_{\sigma}\mathring{u}_{\sigma}\mathring{u}_{\sigma}^+=\frac{1+ \gamma^0}{2}\,,
\quad 	
\sum_{\sigma}	\mathring{v}_{\sigma}\mathring{v}_{\sigma}^+=\frac{1- \gamma^0}{2}\,.
\end{equation}
We have now the opportunity of introducing the Pauli spinors we need for studying the polarization assuming that in the chiral rep. of the Dirac matrices (with diagonal $\gamma^5$) we may express the momentum-dependent spinors as  
\begin{eqnarray}
	\mathring{u}_{\sigma}({\bf p})&=&r({\bf p})	\mathring{u}_{\sigma}=\frac{1}{\sqrt{2}}\left(
	\begin{array}{c}
		\xi_{\sigma}({\bf p})\\
		\xi_{\sigma}({\bf p})
	\end{array}\right)\,,\label{xy}\\
\mathring{v}_{\sigma}({\bf p})&=&r({\bf p})	\mathring{v}_{\sigma}=\frac{1}{\sqrt{2}}\left(
	\begin{array}{c}
		\eta_{\sigma}({\bf p})\\
		-\eta_{\sigma}({\bf p})
	\end{array}\right)\,,\label{xy1}
\end{eqnarray}
in terms of Pauli spinors $\xi_{\sigma}({\bf p})$ and  $\eta_{\sigma}({\bf p})=i\sigma_2 \xi_{\sigma}({\bf p})^*$ that form related bases, i. e. orthonormal,   
\begin{equation}\label{xyort}
\xi_{\sigma}^+({\bf p})\xi_{\sigma'}({\bf p})=\eta_{\sigma}^+({\bf p})\eta_{\sigma'}({\bf p})=\delta_{\sigma\sigma'}\,,	
\end{equation}
and complete systems, 
\begin{equation}\label{xycom}
\sum_{\sigma}\xi_{\sigma}({\bf p})\xi_{\sigma}^+({\bf p})=\sum_{\sigma}\eta_{\sigma}({\bf p})\eta_{\sigma}^+({\bf p})=1_{2\times2}\,.	
\end{equation}
The functions $\xi_{\sigma} : \mathbb{R}^3_p \to {\cal V}_{P}$ remain arbitrary representing the polarization degrees of freedom which will be determined after defining the polarization operators.   As mentioned before, when these spinors depend explicitly on ${\bf p}$  we say that we have  a {\em peculiar} polarization while a  polarization independent on ${\bf p }$ will be referred as {\em common} polarization.  

Finally, by setting the  normalization factor as
\begin{equation}\label{Nor}
	N(p)=\sqrt{\frac{m}{E(p)}}\,,
\end{equation}	
we obtain  the expressions of the spinors (\ref{U}) and (\ref{V}), 
\begin{eqnarray}
	U_{{\bf p},\sigma}(t,{\bf x})&=&N(p)l_{\bf p}\,  \mathring{u}_{\sigma}({\bf p})\frac{1}{(2\pi)^{\frac{3}{2}}} \,e^{-iE(p)t+i{\bf p}\cdot{\bf x}}\,,\label{Ufin} \\
	V_{{\bf p},\sigma}(t,{\bf x})&=&N(p)l_{\bf p} \mathring{v}_{\sigma}({\bf p})\frac{1}{(2\pi)^{\frac{3}{2}}}\, e^{iE(p)t-i{\bf p}\cdot{\bf x}}\,,\label{Vfin}
\end{eqnarray} 
that satisfying Eqs. (\ref{ortU}), (\ref{ortV}) and (\ref{comp}) .   

With these preparations we arrive at the principal result of Wigner's approach showing that the Pauli spinors $\alpha$ and $\beta$ transform alike under  Wigner's  rep. $\tilde T$ acting as \cite{Wig,WKT,Th}
\begin{eqnarray}
	(\tilde T_{\lambda,a}\, \alpha)_{\sigma}({\bf p})=\sqrt{\frac{E(p')}{E( p)}}e^{i {a}\cdot{p}}\sum_{\sigma'}{D}_{\sigma\sigma'}(\lambda,{\bf p}) \alpha_{\sigma'}({{\bf p}}') \,,\label{Wig}
\end{eqnarray}
where the matrix elements 
\begin{equation}\label{Duu}
{D}_{\sigma\sigma'}(\lambda,{\bf p})={\mathring{u}}^+({\bf p})_{\sigma}w(\lambda,{\bf p})\mathring{u}_{\sigma'}({\bf p}')\,,
\end{equation}
depend on the spinors (\ref{xy}) and Wigner transfs. $w(\lambda,{\bf p})=l^{-1}_{{\bf p}} \lambda\, l_{{\bf p}'}$. We observe that the corresponding Lorentz transf.  $\Lambda[w(\lambda,{\bf p})]=L^{-1}_{{\bf p}}\Lambda(\lambda) L_{{{\bf p}}'}$ leaves invariant the representative momentum, as we have $\Lambda[w(\lambda,{\bf p})]\mathring{p}=L^{-1}_{{\bf p}}\Lambda(\lambda){{\bf p}\,}'=L^{-1}_{{\bf p}}{\bf p}=\mathring{p}$,  such that   $\Lambda[w(\lambda,{\bf p})]\in SO(3) \to w(\lambda,{\bf p})\in SU(2)$. Therefore,  we may write the definitive form of the matrix elements (\ref{Duu}) as
\begin{equation}\label{Wrot}
{D}_{\sigma\sigma'}(\lambda,{\bf p})=\xi^+_{\sigma}({\bf p}) \hat l^{-1}_{{\bf p}} \hat\lambda\, \hat l_{{\bf p}'}\xi_{\sigma'}({\bf p}') \,.	
\end{equation}
Obviously, the  matrices $D(\lambda,{\bf p})$ form the irreducible rep. of spin $s=\frac{1}{2}$ of the $SU(2)$ group which means that the Wigner irreducible reps. $\tilde T$ are  {\em  induced} by the subgroup $T(4)\,\circledS\,SU(2)$ \cite{Wig,WKT,Th} as we show  in the Appendix B. For the antiparticle spinors  $\beta^*$ we obtain     the matrix elements 
\begin{eqnarray}
{\mathring{v}}^+_{s\sigma}({\bf p})w(\lambda,{\bf p})\mathring{v}_{s\sigma'}({\bf p}')=\left({\mathring{u}}^+_{\sigma}({\bf p})w(\lambda,{\bf p})\mathring{u}_{\sigma'}({\bf p}')\right)^*
= [{D}_{\sigma\sigma'}(\lambda,{\bf p})]^*  \,,
\end{eqnarray}
showing that  the spinors $\alpha$ and $\beta$  transform alike under isometries.  Note that the form of the matrix elements (\ref{Wrot}) involves explicitly the Pauli spinors helping us to study their dependence on polarization. 

The Wigner rep. $\tilde T$ is irreducible as the matrices ${D}$ are irreducible. Moreover, these are unitary with respect to the scalar product (\ref{spa}) \cite{Wig,Mc}, 
\begin{equation}
	\langle \tilde T_{\lambda,a} \alpha, \tilde T_{\lambda,a} \alpha'\rangle =\langle  \alpha,  \alpha'\rangle	\,,
\end{equation}
and similarly for  $\beta$. As the covariant reps. are unitary with respect to the scalar product (\ref{sp}) which can be decomposed as in Eq. (\ref{spp}) we conclude that the expansion (\ref{Psi}) establishes the unitary equivalence, $T  =\tilde T  \oplus \tilde T $, of the covariant rep. with the {orthogonal} sum  of Wigner's unitary irreducible reps.  \cite{Mc}. This means that the generators $\tilde X\in {\rm Lie}(\tilde T)$   defined as
\begin{eqnarray}
 \tilde	P_{\mu}=-\left.i\frac{\partial \tilde T_{1,a}}{\partial a^{\mu}}\right|_{a=0}\,, \quad 
\tilde	J_{\mu\nu}=\left.i\frac{\partial \tilde T_{\lambda(\omega),0}}{\partial \omega^{\mu\nu}}\right|_{\omega=0}\,,
\end{eqnarray}
are related to the corresponding generators $X\in {\rm Lie}(T)$ such that
\begin{eqnarray}\label{basicx}
	(X\psi)(x)=\int d^3p \sum_{\sigma}\left[U_{{\bf p},\sigma}(x)(\tilde X\, \alpha)_{\sigma}({\bf p})-V_{{\bf p},\sigma}(x) (\tilde X\,\beta)^ {*} _{ \sigma}({\bf p})\right]\,,
\end{eqnarray}
as we deduce deriving Eq. (\ref{basic}) with respect to a group parameter $\zeta\in (\omega, a)$ in $\zeta=0$. 

\section{Looking for spin and coordinate operators}

The next step is to define the polarization looking for operators acting on the space of Pauli spinors ${\cal V}_P$. As  the representative momentum corresponds to a set of rest frames related among themselves through $SO(3)$ rotations of  stable group we observe that the space of Pauli's spinors  has similar  degrees of freedom governed by the $SU(2)$ little group. These degrees of freedom  deserve to be investigated as a symmetry neglected so far.  For this purpose  it is convenient to re-denote  the Dirac field (\ref{Psi}) by $\psi_{\xi}(x)$ and the mode spinors (\ref{Ufin}) and (\ref{Vfin}) by $U_{{\bf p},\xi_{\sigma}}$ and respectively $V_{{\bf p},\eta_{\sigma}}$, pointing out explicitly their dependence on Pauli's spinors. 

On the other hand, we take into account that there are no differential or multiplicative operators acting directly on the Pauli spinors without affecting other quantities. Therefore, these must be more general operators as the integral ones defined by kernels whose Fourier transforms are  usual operators of the ${\bf p}$-rep. used in various applications. In what follows we consider such operators for constructing the spectral reps. we need in our investigation. 

\subsection{Spectral representation of integral  operators}

We focus on the integral operators, $Z: {\cal F}_D\to {\cal F}_D$ whose action, 
\begin{equation}
	(Z\psi)(x)=\int d^4 x' {\cal K}_Z(x-x')\psi(x')\,,
\end{equation}
is determined by the kernels ${\cal K}_Z$ which are  $4\times4$-matrices depending on $x-x'$. These operators are linear forming an algebra in which the multiplication is defined by the convolution, 
\begin{equation}
	{\cal K}_{Z_1Z_2}(x-x')=\int d^4x"{\cal K}_{Z_1}(x-x"){\cal K}_{Z_2}(x"-x')\,,
\end{equation} 
denoted as ${\cal K}_{Z_1Z_2}={\cal K}_{Z_1}*{\cal K}_{Z_2}$. The identity operator  has the kernel ${\cal K}_1(x-x')=\delta^4(x-x')$. An operators $Z$ is invertible if there exists the operator $Z^{-1}$ such that  ${\cal K}_{Z}*{\cal K}_{Z^{-1}}={\cal K}_{Z^{-1}}*{\cal K}_{Z}={\cal K}_{1}$. For any integral operator $Z$ we may write the bracket
\begin{equation}
	\langle \psi,Z\psi'\rangle_D=\int d^4x\, d^4x' \psi^+(x){\cal K}_Z(x-x')\psi(x')
\end{equation} 
observing that $Z$ is self-adjoint with respect to this scalar product only if ${\cal K}_Z(x)={\cal K}_Z^+(-x)$. Note that the multiplicative or differential operators are particular cases of integral ones. For example, the derivatives $\partial_{\mu}$ can be seen as  integral operators having the kernels ${\cal K}_{\partial_{\mu}}(x)=\partial_{\mu}\delta^4(x)$. 

Of a  special interest are  the equal-time operators $Y$ whose kernels have the form 
\begin{equation}
	{\cal K}_Y(x-x')=\delta(t-t'){ \cal K}_Y({\bf x}-{\bf x}')\,,
\end{equation}  
acting as
\begin{equation}\label{Y1}
	(Y \psi)(t,{\bf x})=\int d^3x' {\cal K}_Y({\bf x}-{\bf x}')\psi(t,{\bf x}')\,,
\end{equation}
without involving the time. In this case the kernels  allow the $3$-dimensional Fourier rep., 
\begin{equation}\label{KerY}
	{\cal K}_{Y}({\bf x}) =\int d^3p\,\frac{e^{i {\bf p}\cdot{\bf x}}}{(2\pi)^3} \tilde {Y}({\bf p})\,,
\end{equation} 
where $\tilde Y({\bf p})$ is the operator in ${\bf p}$-rep. of RQM corresponding to $Y$. Then the action (\ref{Y1}) on a field  (\ref{Psi})  can be written as
\begin{eqnarray}
(Y \psi_{\xi})( x)=\int d^3p \sum_{\sigma}\left[ \tilde Y({\bf  p})U_{{\bf p},\xi_{\sigma}}( x)\alpha_{\sigma}({\bf p}) +\tilde Y(-{\bf  p})V_{{\bf p},\eta_{\sigma}}(x)\beta^*_{\sigma}({\bf p})\right]\,.\label{Y2}
\end{eqnarray}
We see thus that all these integral operators are diagonal on ${\cal F}_D$, acting separately on ${\cal F}_D^{\pm}$ without mixing mode spinors of different frequencies.

In what follows we consider a special type of such equal-time operators, denoted by $Y_{\xi}$, acting alike on the spinors $\alpha$ and $\beta$, defined by kernels that allow the spectral rep.
\begin{eqnarray}\label{kerY}
	{\cal K}_{Y_{\xi}}({\bf x}-{\bf x}')&=&\int d^3p\sum_{\sigma,\sigma'}\left[U_{{\bf p},\xi_{\sigma}}(t,{\bf x})\tilde y_{\sigma\sigma'}U^+_{{\bf p},\xi_{\sigma'}}(t,{\bf x}')\right.\nonumber\\
	&&\hspace*{20mm}\left. +V_{{\bf p},\eta_{\sigma}}(t,{\bf x})\tilde y^{*}_{\sigma\sigma'}V^+_{{\bf p},\eta_{\sigma'}}(t,{\bf x}')\right]\,,
\end{eqnarray} 
in terms of mode spinors (\ref{Ufin}) and (\ref{Vfin}) depending on arbitrary Pauli spinors $\xi \subset {\cal V}_P$ that may depend on ${\bf p}$. The action of these operators  can be calculated easily  in ${\bf p}$-rep.  by using   the orthogonality properties  (\ref{ortU}) and  (\ref{ortV}).  Then it turns out the action in ${\bf x}$-rep.,
\begin{eqnarray}\label{actY}
	(Y_{\xi}\psi_{\xi})(x)&=&\int d^3 x' {\cal K}_{Y_{\xi}}({\bf x}-{\bf x}') \psi_{\xi}(t,{\bf x}')\nonumber\\
&=&\int d^3p\sum_{\sigma,\sigma'}\left[U_{{\bf p},\xi_{\sigma}}(t,{\bf x})\tilde y_{\sigma\sigma'}\alpha_{\sigma'}({\bf p}) +V_{{\bf p},\eta_{\sigma}}(t,{\bf x})\tilde y^{*}_{\sigma\sigma'} \beta^*_{\sigma'}({\bf p})\right]\,,~~~
\end{eqnarray}
related to the action of  associated matrix-operator $\tilde y({\bf p})$ transforming alike the  spinors $\alpha$ and $\beta$,
\begin{eqnarray}
	(\tilde y\alpha)_{\sigma}({\bf p})&=&\langle U_{{\bf p},\xi_{\sigma}}, Y_{\xi}\psi_{\xi}\rangle_D=\tilde y_{\sigma\sigma'}\alpha_{\sigma'}({\bf p})\,,\\
		(\tilde y \beta)_{\sigma}({\bf p})&=&\langle Y_{\xi}\psi_{\xi}, V_{{\bf p},\eta_{\sigma}}\rangle_D=\tilde y_{\sigma\sigma'}\beta_{\sigma'}({\bf p})\,.
\end{eqnarray}
The special form of these operators allow us to derive their expectation values, 
\begin{equation}\label{clou}
	\langle \psi_{\xi}, Y_{\xi}\psi_{\xi}\rangle_D=\langle \alpha, \tilde y\,\alpha\rangle +\langle \beta, \tilde y^+\beta\rangle\,, 
\end{equation}
exploiting Eqs. (\ref{ortU}) and  (\ref{ortV}).  Hereby we see that an operator $Y_{\xi}$ is self-adjoint if the matrix $\tilde y$ is Hermitian, $\tilde y_{\sigma\sigma'}=\tilde y_{\sigma'\sigma}^*$. 

We must stress that the dependence on $\xi$ of the operator $Y_{\xi}$ is not an impediment as we know what happens when we change the spinor basis. Thus, by using Eqs. (\ref{Uxi}) and (\ref{Veta}) we find that $Y_{\xi}\to	Y_{\hat r \xi}~ \Rightarrow ~\tilde y \to D(\hat r)\tilde y$ if we keep unchanged the spinors $\alpha$ and $\beta$ encapsulating the physical meaning.

\subsection{Spin and polarization}

For exploiting the spin degrees of freedom we start with an arbitrary orthonormal  basis $\xi \subset {\cal V}_P$, satisfying Eqs. (\ref{xyort}) and (\ref{xycom}), whose spinors may depend on ${\bf p}$ but without denoting this explicitly.  The rotations $\hat r \in SU(2)$ of the little group transform this  basis as
\begin{eqnarray}
	\hat r\, \xi_{\sigma}&=&\sum_{\sigma'}\xi_{\sigma'}D_{\sigma'\sigma}(\hat r)   \Rightarrow r\, \mathring{u}_{\sigma}= \sum_{\sigma'}\mathring{u}_{\sigma'}D_{\sigma'\sigma}(\hat r)  \,,\label{rotx}\\
	\hat r\, \eta_{\sigma}&=&\sum_{\sigma'}\eta_{\sigma'}D^*_{\sigma'\sigma}(\hat r)  \Rightarrow r\, \mathring{v}_{\sigma}=\sum_{\sigma'}\mathring{v}_{\sigma'}D^*_{\sigma'\sigma}(\hat r)\,,\label{roty}
\end{eqnarray}
where $r={\rm diag} (\hat r, \hat r)\in \rho_D$ is an arbitrary  rotation  corresponding to $\hat r$  for which we use the traditional notation  
\begin{equation}\label{Dr}
	D_{\sigma'\sigma} (\hat r)=	\xi_{\sigma'}^+\hat r \xi_{\sigma}\,.
\end{equation}
 We obtain thus the transfs. of mode spinors,  
\begin{eqnarray}
	U_{{\bf p},\hat r \xi_{\sigma}}(x)&=&\sum_{\sigma'} U_{{\bf p},\xi_{\sigma'}}(x)D_{\sigma'\sigma}(\hat r)\,,\label{Uxi}\\
	V_{{\bf p},\hat r \eta_{\sigma}}(x)&=&\sum_{\sigma'} V_{{\bf p},\eta_{\sigma'}}(x)D^*_{\sigma'\sigma}(\hat r)\,,\label{Veta}
\end{eqnarray}
which give the transformed free field $\psi_{\hat r\xi}$ according to the expansion (\ref{Psi}). 

Under such circumstances, we may look for  a rep.,  ${\frak R}: \hat r\to {\frak R}(\hat r)$, of the little group $SU(2)$ with values in a set of operators  ${\frak S}=\{{\frak R}(\hat r)| \hat r\in SU(2)\}$ able to rotate the Pauli spinors but without depending on the spinor basis $\xi$ or affecting other quantities. These operators  can be constructed as integral  operators with kernels of the form (\ref{kerY}) where 
\begin{equation}\label{yDxx}
\tilde y=D(\hat r)~\Rightarrow~\tilde y_{\sigma\sigma'}=D_{\sigma\sigma'}(\hat r)=\xi^+_{\sigma}\hat r \xi_{\sigma'}\,.	
\end{equation}
By substituting then these matrices in Eq. (\ref{actY}) and applying the identities (\ref{Uxi}) and (\ref{Veta}) we find the desired action  
\begin{eqnarray}
	[{\frak R}(\hat r)\psi_{\xi}](t,{\bf x})=\int d^3x' {\cal K}_{{\frak R}(\hat r)}({\bf x}-{\bf x}')\psi_{\xi}(t,{\bf x}')	=\psi_{\hat r\xi}(t,{\bf x})\,,
\end{eqnarray}
upon the basis of Pauli spinors. In addition, the general rule (\ref{clou}) allows us to derive the expectation values of these operators
\begin{equation}\label{exV}
	\langle \psi_{\xi},	{\frak R}(\hat r)\psi_{\xi}\rangle_D=\langle \alpha, D(\hat r) \alpha\rangle +\langle \beta, D^+(\hat r ) \beta\rangle\,,
\end{equation}
which depend explicitly on $\xi$ through the matrix $D(\hat r)$. 

Furthermore, we consider the mode spinors (\ref{Ufin}) and (\ref{Vfin}),  the action of $SU(2)$ rotations (\ref{rotx}) and (\ref{roty}) as well the identities (\ref{idll}) and (\ref{idlll}),  deducing that the integral operators ${\frak R}(\hat r)\in {\frak S}$ have
kernels of the form (\ref{KerY}) whose  Fourier transforms read    
\begin{eqnarray}
\tilde {\frak R}({\hat r,\bf p})&=&\frac{m}{E(p)}\left[ l_{\bf p}\,  r\, \frac{1+\gamma^0}{2}l_{\bf p}+l^{-1}_{\bf p} r\, \frac{1-\gamma^0}{2}l^{-1}_{\bf p}\right] \nonumber\\
&=& l_{\bf p}  r\,  l_{\bf p}^{-1}\tilde \Pi_+({\bf p}) + l_{\bf p}^{-1}  r\,  l_{\bf p}\tilde \Pi_-({\bf p})	\,,
\end{eqnarray}
where $r={\rm diag}(\hat r,\hat r)\in\rho_D$ while $\tilde \Pi_{\pm}({\bf p})$ are the projection operators (\ref{Pip}) and (\ref{Pim}). The operator ${\frak R}(\hat r)$ is independent on the spinor bases under consideration as  the spinors $\mathring{u}({\bf p})$ and $\mathring{v}({\bf p})$  satisfy similar relations as (\ref{ortuv}) because  $r({\bf p})$ commutes with $\gamma^0$.

We defined thus the set ${\frak S}$ of operators whose properties can be studied in  ${\bf p}$-rep. as their   Fourier transforms obey the same algebra,  
\begin{equation}
{\frak R}={\frak R}_1{\frak R}_2~\Rightarrow~{\cal K}_{\frak R}={\cal K}_{{\frak R}_1}*{\cal K}_{{\frak R}_2}~\Rightarrow~\tilde {\frak R}({\bf p})=\tilde {\frak R}_1({\bf p}) \tilde {\frak R}_2({\bf p})\,.
\end{equation} 
 By using then the identities (\ref{idll}) and (\ref{idlll}), after a little calculation,  we verify that
\begin{equation}
\tilde {\frak R}(\hat r,{\bf p})\tilde {\frak R}(\hat r', {\bf p})=\tilde {\frak R}(\hat r\hat r',{\bf p})\,,
\end{equation}
observing that for $\hat r=1_{2\times 2}$  we have   
\begin{equation}
\tilde{\frak R}(1_{2\times2},{\bf p})=\tilde\Pi_+({\bf p})+\tilde\Pi_-({\bf p}) =1\in \rho_D  \,.
\end{equation}
We conclude that the set  ${\frak S}$ form just the  $SU(2)$ rep. we are looking for. We  arrive thus at our principal objective, namely the definition of spin operator.

{\bf Definition 1}
{\em  The spin operator is the vector-operator ${\bf S}$ whose components form a canonical basis of the algebra  ${\rm Lie}({\frak R})\sim su(2)$.} \\
 Starting with the transf. ${\frak R}(\hat r(\theta))$ depending on the rotation (\ref{r}) we derive the spin comps., 
 \begin{equation}
 	S_i=\left.i\frac{\partial {\frak R}(\hat r(\theta))}{\partial \theta^i}\right|_{\theta^i=0}\,,
 \end{equation}
finding that they are integral operators acting as
 \begin{eqnarray}
 	[S_i\psi_{\xi}](t,{\bf x})&=&\int d^3x' {\cal K}_{S_i}({\bf x}-{\bf x}')\psi_{\xi}(t,{\bf x}')\nonumber\\
 	&=&\psi_{\hat s_i\xi}(t,{\bf x})\,,
 \end{eqnarray}
through  kernels having as  Fourier transforms the spin comps. in ${\bf p}$-rep.,    
\begin{eqnarray}\label{Sip}
	\tilde S_i({\bf p})&=&\frac{m}{E(p)}\left[ l_{\bf p}\, s_i\, \frac{1+\gamma^0}{2}l_{\bf p}+l^{-1}_{\bf p}s_i\, \frac{1-\gamma^0}{2}l^{-1}_{\bf p}\right] \nonumber\\
	&=&s_i({\bf p}) \tilde \Pi_+({\bf p}) +  s_i ( -{\bf p})\tilde \Pi_-({\bf p})	\,,	
\end{eqnarray}
where $s_i({\bf p})= l_{\bf p}  s_i  l_{\bf p}^{-1}$  are the comps. of the transformed reducible Pauli spin operator. In the rest frame we have ${\bf p}=0 \Rightarrow \tilde{\bf S}(0)={\bf s}(0)={\bf s}$. 

Surprisingly, after a little calculation, we find that the operators (\ref{Sip}) are just the comps. of Pryce's spin operator, 
\begin{equation}\label{final}
	\tilde S_i({\bf p})=\frac{m}{E(p)} s_i+\frac{p^i ({\bf s}\cdot{\bf p})}{E(p)(E(p)+m)} +\frac{i}{2E(p)}\epsilon_{ijk}p^j \gamma^k \,,
\end{equation}
found long time ago  (see the third of Eqs. (6.7) of Ref. \cite{B}) in association with a would be relativistic mass-center coordinate operator. The same spin operator was defined alternatively as in Eq. (\ref{FW3}) such that it becomes the Pauli one in the frame where the Hamiltonian is $\gamma^0E(p)$ instead of the rest frame as in the case or our definition. However, the principal novelty of our definition is of pointing out that the  spin  comps. are the generators of a  Noetherian symmetry. Thus we conclude that Def. 1 is different from those of Pryce or Foldy and Wouthuysen.

By definition, the operators (\ref{Sip}) generate  the $su(2)$ algebra,  
\begin{equation}
	\left[	\tilde S_i({\bf p})  ,	\tilde S_j({\bf p})\right]=i\epsilon_{ijk}	\tilde S_k({\bf p})~\Rightarrow ~
	\left[ S_i  ,	 S_j\right]=i\epsilon_{ijk} S_k\,,
\end{equation}
are self-adjoint and conserved commuting with the Dirac Hamiltonian in ${\bf p}$-rep. (\ref{HDp}).  The action of these operators on the mode spinors (\ref{Ufin}) and (\ref{Vfin}) can be derived as in Eq. (\ref{Y2}) obtaining
\begin{eqnarray}
	(S_i U_{{\bf p},\xi_{\sigma}})(x)&=&\tilde S_i({\bf p})U_{{\bf p},\xi_{\sigma}}(x)=U_{{\bf p},\hat s_i\xi_{\sigma}}(x)\,,\\
	(S_i V_{{\bf p},\eta_{\sigma}})(x)&=&\tilde S_i(-{\bf p})V_{{\bf p},\eta_{\sigma}}(x)=V_{{\bf p},\hat s_i\eta_{\sigma}}(x)\,,	
\end{eqnarray}
after using the form (\ref{Sip}) and  the identities (\ref{idll}) and (\ref{idlll}). 

In other respects, the polarization is given by the pair of related spinors $\xi_{\sigma}({\bf p})$  and $\eta_{\sigma}({\bf p})$ assumed to satisfy the eigenvalues problems
\begin{equation}\label{snpp}
\hat s_i  {n}^i({\bf p})\xi_{\sigma}({\bf p})	=\sigma\, \xi_{\sigma}({\bf p}) \to 
\hat s_i  {n}^i({\bf p})\eta_{\sigma}({\bf p})	=-\sigma\, \eta_{\sigma}({\bf p}),
\end{equation}
where the unit vector ${\bf n}({\bf p})$ give the peculiar direction with respect to which the peculiar polarization is measured.  Under such circumstances we may define a convenient polarization operator appropriate to the present framework.

{\bf Definition 2}
{\em The polarization operator is the integral operator $W$ whose kernel has the Fourier transform
\begin{eqnarray}
	\tilde W({\bf p})=w({\bf p})\Pi_+({\bf p}) +  w(-{\bf p})\tilde \Pi_-({\bf p})
	\,,	\label{Pol}
\end{eqnarray}
where $w({\bf p})= l_{\bf p}  s_i n^i({\bf p}) l_{\bf p}^{-1}$.}	

This operator is conserved, commuting with the Hamiltonian operator (\ref{HDp}). Moreover, it commutes with $P^i$ and $H$ acting on the mode spinors constructed with the spinors of Eqs. (\ref{snpp}) as 
\begin{eqnarray}
	(W U_{{\bf p},\xi_{\sigma}({\bf p})})(x)=\tilde W({\bf p})U_{{\bf p},\xi_{\sigma}({\bf p})}(x)&=&U_{{\bf p},\hat s_i n^i({\bf p})\xi_{\sigma}({\bf p})}(x)\nonumber\\
	&=&\sigma U_{{\bf p},\xi_{\sigma}({\bf p})}(x)\,,\label{WU}\\
	(W V_{{\bf p},\eta_{\sigma}({\bf p})})(x)=\tilde W(-{\bf p})V_{{\bf p},\eta_{\sigma}({\bf p})}(x)&=&V_{{\bf p},\hat s_i  n^i({\bf p}) \eta_{\sigma}({\bf p})}(x)\nonumber\\
	&=& -\sigma V_{{\bf p},\eta_{\sigma}({\bf p})}(x)\,.\label{WV}	
\end{eqnarray}
These eigenvalues problems convince us that $W$ is just the operator we need for completing the system of commuting operators as	$\{H,P^1,P^2,P^3, W\}$ for defining properly the ${\bf p}$-reps. of RQM.  

As the comps. of spin operator are integral operators we understand that $W$ remains an operator of this type even in the case of common polarization when ${\bf n}$ and $\xi$ are independent on ${\bf p}$ and, consequently, we may write $W={\bf S}\cdot{\bf n}$. The well-known example is the momentum-spin basis \cite{BDR} where ${\bf n} ={\bf e}_3$ while $W=S_3$ defines the basis spinors  
\begin{equation}\label{etapm}
	\xi_{\frac{1}{2}}=\left(\begin{array}{c}
		1\\
		0
	\end{array}\right)\,,\quad
	\xi_{-\frac{1}{2}}=\left(\begin{array}{c}
		0\\
		1
	\end{array}\right)  \,.
\end{equation} 
used in various applications \cite{W,BDR,KH}. Therefore, we may say that there are no situations in which the polarization might be defined properly by a differential polarization operator  as, for example, the helicity one, $W_0$.  We shall discuss later the differences between these two operator when we will  study the polarization and spin operators in momentum-helicity basis.  

\subsection{Associated coordinate operator}

 Looking for a mass-center position vector, Pryce assumed that this has the form  ${\bf X}=\underline{\bf x}+\delta{\bf X}$ focusing on the correction $\delta{\bf X}$ which, according to the identity (\ref{spli}), must satisfy $\delta{\bf X}\land {\bf P}+{\bf S}={\bf s}$.  Analyzing various hypotheses,  Pryce concluded that $\delta{\bf X}$ is an integral operator whose comps. have kernels given by the Fourier transforms \cite{B}
\begin{equation}
\delta \tilde X^i({\bf p})	=\frac{i\gamma^i}{2E(p)}+\frac{\epsilon_{ijk}p^js_k}{E(p)(E(p)+m)} -\frac{ip^i \gamma^jp^j}{2E(p)^2(E(p)+m)}\,,
\end{equation}  
which satisfies the desired  identity $\delta \tilde {\bf X}({\bf p})\land{\bf p}+\tilde{\bf S}({\bf p})={\bf s}$. These are self-adjoint operators commuting with ${\bf P}$ such that 
\begin{equation}\label{XPij}
	[X^i, P^j]=[\underline{x}^i,P^j]=i \delta_{ij} 1 \in \rho_D\,.
\end{equation} 
Other properties including commutation relations have to be studied after quantization in order to avoid tedious calculations in ${\bf p}$-rep..

By using then suitable codes on computer it is not difficult to verify that these operators can be put in the form 
\begin{eqnarray}
\delta \tilde X^i({\bf p})=\delta x^i({\bf p})\tilde\Pi_+({\bf p}) +  \delta x^i(-{\bf p})\tilde \Pi_-({\bf p})\,,
\end{eqnarray}
where 
\begin{equation}
	\delta x^i({\bf p})=-i\frac{1}{N(p)}\partial_{p^i} \left(N(p)l_{\bf p}\right)l^{-1}_{\bf p}\,.
\end{equation}
Furthermore, observing that these operators derive only the factor $N(p) l_{\bf p}$ of the mode spinors (\ref{Ufin}) and (\ref{Vfin})  we may write their action as
\begin{eqnarray}
\left(\delta X^i U_{{\bf p},\xi_{\sigma}}\right)(t,{\bf x})&=&\delta \tilde X^i({\bf p})U_{{\bf p},\xi_{\sigma}}(t,{\bf x})	=-i\partial_{p^i}U_{{\bf p},\xi_{\sigma}}(t,{\bf x})\nonumber\\
&&-x^i U_{{\bf p},\xi_{\sigma}}(t,{\bf x})+\frac{t p^i}{E(p)}U_{{\bf p},\xi_{\sigma}}(t,{\bf x})\nonumber\\
&&+\sum_{\sigma'}U_{{\bf p},\xi_{\sigma'}}(t,{\bf x})\Omega_{i\,\sigma' \sigma}({\bf p})\,,\label{X1}\\
\left(\delta X^i V_{{\bf p},\eta_{\sigma}}\right)(t,{\bf x})&=&\delta \tilde X^i(-{\bf p})V_{{\bf p},\eta_{\sigma}}(t,{\bf x})	=i\partial_{p^i}V_{{\bf p},\eta_{\sigma}}(t,{\bf x})\nonumber\\
&&-x^i V_{{\bf p},\eta_{\sigma}}(t,{\bf x})+\frac{t p^i}{E(p)}V_{{\bf p},\eta_{\sigma}}(t,{\bf x})\nonumber\\
&&-\sum_{\sigma'}V_{{\bf p},\eta_{\sigma'}}(t,{\bf x})\Omega^*_{i\,\sigma' \sigma}({\bf p})\,.\label{X2}
\end{eqnarray}
Here we use the artifice
\begin{equation}
	\partial_{p^i}\xi_{\sigma}({\bf p})=\sum_{\sigma'}\xi_{\sigma'}({\bf p})\Omega_{i\,\sigma',\sigma}({\bf p})\,,
\end{equation}
and similarly for the spinors $\eta_{\sigma}({\bf p})$, denoting  
\begin{equation}\label{Omega}
	\Omega_{i\,\sigma\sigma'}({\bf p})=\xi^+_{\sigma}({\bf p})\left[\partial_{p^i}\xi_{\sigma'}({\bf p})\right]=\left\{\eta^+_{\sigma}({\bf p})\left[\partial_{p^i}\eta_{\sigma'}({\bf p})\right]\right\}^*\,,
\end{equation}
and observing that $\Omega_{i\,\sigma\sigma'}({\bf p})=-\Omega_{i\,\sigma'\sigma}^*({\bf p})$ which means that the matrices $i\Omega_i$ are Hermitian. 

Hereby we understand that the Pryce coordinate operator depends linearly on time  as ${\bf X}(t)=\underline{\bf x} + \delta {\bf X}(t)={\bf X}+{\bf V}t$. From Eqs. (\ref{X1}) and (\ref{X2}) and applying the Green theorem in the integral over momenta we find  the actions of these operators, 
\begin{eqnarray}
(X^i\psi_{\xi})(x)&=&\int d^3p \sum_{\sigma}\left[U_{{\bf p},\xi_{\sigma}}(x)	i\tilde\partial_{i}\alpha_{\sigma}({\bf p})-V_{{\bf p},\xi_{\sigma}}(x)i\tilde\partial_{i} \beta^*_{\sigma'}({\bf p})\right]\,,\label{ActX}\\
(V^i\psi_{\xi})(x)&=&\int d^3p \frac{p^i}{E(p)} \sum_{\sigma}\left[U_{{\bf p},\xi_{\sigma}}(x) \alpha_{\sigma}({\bf p})  +V_{{\bf p},\eta_{\sigma}}(x) \beta^*_{\sigma}({\bf p})\right]\,,\label{ActV}
\end{eqnarray} 
written in terms of the 'covariant' derivatives 
\begin{equation}\label{covd}
	\tilde\partial_i \alpha_{\sigma}({\bf p})=\partial_{p^i}\alpha_{\sigma}({\bf p})+\sum_{\sigma'}	\Omega_{i\,\sigma\sigma'}({\bf p})\alpha_{\sigma'}({\bf p})\,,
\end{equation}
where the matrices $\Omega_i({\bf p})$ play the role of connection. These derivatives commute among themselves, $[\tilde\partial_i, \tilde\partial_j]=0$, assuring that  $[X^i,X^j]=0$. 

The operator ${\bf X}$ is the Pryce coordinate operator at $t=0$ while ${\bf V}$ is a conserved velocity. Bearing in mind that in RQM the antiparticle terms cannot be interpreted properly, we may ask if the Pryce assumption that these kinetic quantities describe the inertial motion of mass-center is correct or not.   We shall find the answer after performing the  quantization.  

\section{One-particle operators of quantum theory}

The principal benefit of our approach based on spectral reps. is the relation between the operator actions  in ${\bf x}$  and ${\bf p}$-reps. allowing us to derive at any time the expectation values of  operators defined in ${\bf p}$-rep..  We get thus the opportunity of applying the Bogolyubov method for quantizing the Pryce operators and deriving   the isometry generators of the massive Dirac fermions of  arbitrary polarization.   

\subsection{Quantization and diagonal operators}

Adopting here the Bogolyubov method of quantization \cite{Bog} we replace first  the functions in ${\bf p}$-rep. with field operators, $(\alpha, \alpha^*)\to ({\frak a},{\frak a}^{\dag})$ and $(\beta, \beta^*)\to ({\frak b},{\frak b}^{\dag})$,  satisfying  canonical   anti-commutation relations among them  the non-vanishing ones are, 
\begin{eqnarray}\label{cac}
	\left\{{\frak a}_{\sigma}({\bf p}),{\frak a}_{\sigma'}^{\dag}({\bf p}')\right\}=	\left\{{\frak b}_{\sigma}({\bf p}),{\frak b}_{\sigma'}^{\dag}({\bf p}')\right\}=\delta_{\sigma\sigma'}\delta^3({\bf p}-{\bf p}')\,.
\end{eqnarray}
The Dirac free field $\psi$ (written from now without the index  $\xi$) becomes thus a field operator while the  expectation value of any time-dependent operator $A(t)$  becomes the one-particle operator, 
\begin{equation}
	A(t)~\to ~ \mathsf{A}=\left.:\langle\psi ,A(t)\psi\rangle_D :\right| _{t=0}\,,
\end{equation}
calculated respecting the normal ordering of the operator products \cite{BDR} at the initial time $t=0$ when we assume that the quantization is performed.  

We obtain thus a basis of  operator algebra  formed by field  and one-particle operators which have the obvious properties
\begin{equation}\label{algXX}
	\left[\mathsf{A}, \psi(x)\right]=-(A\psi)(x)\,, \quad
	\left[\mathsf{A}, \mathsf{B}\right]=:\left<\psi, [A,B]\psi\right>_D: \,,
\end{equation} 
preserving the structures of Lie algebras. Note that the quantization does not take over other algebraic properties from RQM as, in general, the product of two one-particle operators is no longer an operator of this type.

The quantization reveals the physical meaning of the quantum observables of RQM transforming them in the one-particle operators of QFT. The simplest example is the identity operator  $1\in\rho_D$ of RQM,  i. e. the generator of the gauge group $U(1)_{em}$, becoming  through quantization the conserved charge operator,
\begin{eqnarray}
	\mathsf{Q}=:\langle\psi, \psi\rangle_D:=\mathsf{Q}_++\mathsf{Q}_-
	=\int d^3p\,\sum_{\sigma}\left[{\frak a}^{\dag}_{\sigma}({\bf p}){\frak a}_{\sigma}({\bf p}) -{\frak b}^{\dag}_{\sigma}({\bf p}){\frak b}_{\sigma}({\bf p})\right]\,,	
\end{eqnarray}	 
where the particle and antiparticle charge operators are  given just by the projection operators (\ref{Pip}) and (\ref{Pim}) as 
\begin{equation}
	\mathsf{Q}_{\pm}= :\langle \psi, \Pi_{\pm}\psi\rangle_D:\,.	
\end{equation}
Then the operator of number of particles, 
\begin{equation}
	\mathsf{N}=\mathsf{Q}_+-\mathsf {Q}_-=:\langle \psi, (\Pi_{+}-\Pi_-)\psi\rangle_D:,	
\end{equation}
is related to the operator whose kernel has the Fourier transform $E(p)^{-1}\tilde H_D({\bf p})$.

In the basis in which the commuting operators  $\{ H, P^1,P^2,P^3,W\}$ are diagonal  we derive  the corresponding one-particle operators,  
\begin{eqnarray}
	\mathsf{H}&=&:\langle\psi, H\psi\rangle_D:=\int d^3p\,E(p)\sum_{\sigma}\left[{\frak a}^{\dag}_{\sigma}({\bf p}){\frak a}_{\sigma}({\bf p}) +{\frak b}^{\dag}_{\sigma}({\bf p}){\frak b}_{\sigma}({\bf p})\right]\,,\label{Hom}\\ 
	\mathsf{P}^i&=&:\langle\psi, P^i\psi\rangle_D:=\int d^3p\,p^i\sum_{\sigma}\left[{\frak a}^{\dag}_{\sigma}({\bf p}){\frak a}_{\sigma}({\bf p}) +{\frak b}^{\dag}_{\sigma}({\bf p}){\frak b}_{\sigma}({\bf p})\right]\,, \label{Pom}	\\	
	\mathsf{W}& =&:\langle \psi, W\psi\rangle_D:=\int d^3p\sum_{\sigma=\pm\frac{1}{2}} \sigma\left[{\frak a}^{\dag}_{\sigma}({\bf p}){\frak a}_{\sigma}({\bf p}) +{\frak b}^{\dag}_{\sigma}({\bf p}){\frak b}_{\sigma}({\bf p})\right]\,, \label{Polq}
\end{eqnarray}
which commute among themselves and with $\mathsf{Q}$. We obtain thus  the complete system  $\{\mathsf{H},\mathsf{P}^1,\mathsf{P}^2, \mathsf{P}^3,\mathsf{W},\mathsf{Q}\}$ determining the bases of the Fock state space. 

The Energy operator $\mathsf{H}$ generates the unitary operators of  time translations giving the operators at any time, 
\begin{equation}
\mathsf{U}(t)=e^{i\mathsf{H} t}: \mathsf{A}~\to~\mathsf{A}(t)=\mathsf{U}(t)\mathsf{A}\mathsf{U}^{\dag}(t) \,,	
\end{equation} 
apart from the {\em  conserved} ones which commute with $\mathsf{H}$. 

\subsection{Spin operator and $SL(2,\mathbb{C})$ generators}

For applying the same method to  Pryce's spin operator  we look for expectation values of its comps. $S_i$ at the level of RQM. These can be found deriving with respect to Cayley-Klein parameters $\theta^i$ the expectation values (\ref{exV}) where we substitute $D(\hat r)\to D[\hat r(\theta)]$ with $\hat r(\theta)$  given by Eq. (\ref{r}). As the  quantization changes the wrong sign of the anti-particle term we obtain the  comps. of the spin operator,  
\begin{eqnarray}
\mathsf{S}_i =:\langle \psi, S_i\psi\rangle_D: =\frac{1}{2}\int d^3p\sum_{\sigma,\sigma'} \Sigma_{i\,\sigma\sigma'}({\bf p})\left[{\frak a}^{\dag}_{\sigma}({\bf p}){\frak a}_{\sigma'}({\bf p})+{\frak b}^{\dag}_{\sigma}({\bf p}){\frak b}_{\sigma'}({\bf p})\right]\,,
 \label{Spin}
\end{eqnarray} 
where  we denoted by  
\begin{equation}\label{Dxx}
 \Sigma_{i\,\sigma\sigma'}({\bf p})=\xi^+_{\sigma}({\bf p})\sigma_i\,\xi_{\sigma'}({\bf p})\,,	
\end{equation}
the matrix elements of Pauli's operators in the $\xi$-basis in which the quantum field $\psi$ was defined.  The operators $\mathsf{S}_i$ are self-adjoint and form the canonical basis of an operator valued rep. of the $su(2)\sim so(3)$ algebra. 

Let us see now how the spin operator defined above is involved in the structure of the $SL(2,\mathbb{C})$  generators. We start with the expectation values,
\begin{equation}\label{XXX1}
	\langle\psi, X\psi\rangle_D=\langle\alpha, \tilde X\alpha\rangle-\langle\beta, \tilde X\beta\rangle\,,
\end{equation}
derived from Eq. (\ref{basicx}) for any pair of related  generators $X\in{\rm Lie}(T)$ and $\tilde X \in{\rm Lie}(\tilde T)$ corresponding to the same group parameter.  Applying then the quantization we change the wrong relative sign ($-$) in Eq.  (\ref{XXX1}) after restoring the normal ordering of operator products, obtaining  correct forms of one-particle operators. The $SL(2, \mathbb{C})$ generators may be derived by using our parametrizations (\ref{r}) and (\ref{l}). 

For deriving the rotation generators we take $\hat\lambda=\hat r(\theta)$ observing that the transformed momentum (\ref{pLp}) can be expanded now as  $p^{\prime\,i}=p^i +\epsilon_{ijk}p^j\theta^k+\cdots$,  according to the general rule (\ref{Lala}). Introducing these quantities  in   Eq. (\ref{Wrot}) and deriving the transf.    (\ref{Wig}) with respect to the Cayley-Klein parameters $\theta^i$ in $\theta^i=0$  it turns out  
\begin{equation}\label{split}
\mathsf{J}_i=:\langle\psi, J_i\psi\rangle_D:=\mathsf{L}_i+\mathsf{S}_i\,,
\end{equation} 
where  the comps. $\mathsf{S}_i$ of the spin operator  are defined by Eq. (\ref{Spin}). The associated orbital angular momentum operator has the comps. 
\begin{eqnarray}\label{Lang}
\mathsf{L}_i =-{i}\int d^3p\, \epsilon_{ijk} p^j \sum_{\sigma}\left[{\frak a}^{\dag}_{\sigma}({\bf p}){\tilde\partial}_{k}{\frak a}_{\sigma}({\bf p})+{\frak b}^{\dag}_{\sigma}({\bf p}){\tilde\partial}_{k}{\frak b}_{\sigma}({\bf p})\right]\,,
\end{eqnarray}
where the derivatives $\tilde\partial_i$ are defined by Eq. (\ref{covd}).
 
The commutation relations of these operators can be derived easily by using Eqs. (\ref{xyort}),  (\ref{xycom}), identities of the form
\begin{equation}\label{idimidi}
\tilde\partial_i{\frak a}_{\sigma}({\bf p})=\sum_{\sigma'}\xi^+_{\sigma}({\bf p})\partial_{p^i}\left[\xi_{\sigma'}({\bf p}){\frak a}_{\sigma'}({\bf p})\right]\,,
\end{equation}
and taking into account that $[\tilde\partial_i,\Sigma_i]=0$.
We find thus that the operators $\mathsf{L}_i $ and  $\mathsf{S}_i$ form the bases of two {\em independent}  $su(2)\sim so(3)$ algebras commuting each other,  $\left[\mathsf{L}_i,\mathsf{S}_j\right]=0$. Moreover, these operators are {\em conserved} separately, each one commuting with the Hamiltonian operator,
\begin{equation}
	\left[\mathsf{H},\mathsf{L}_i\right]=0\,,\qquad
	\left[\mathsf{H},\mathsf{S}_i\right]=0\,.
\end{equation}
 Therefore, we may conclude that the Pryce spin operator of  RQM gives just the conserved one-particles spin  operator we need in QFT.

The generators of the Lorentz boosts can be found by choosing $\hat \lambda=\hat l(\tau)$ as in Eq. (\ref{l}),  observing  that now $p^{\prime \, i} =p^i+\tau^i E(p)+\cdots$ and deriving Eq.  (\ref{Wig}) with respect to $\tau^i$ in $\tau=0$. After a few manipulation we  derive the operators at the initial time $t=0$,  
\begin{eqnarray}
	\mathsf{K}_i &	=&:\langle\psi, K_i\psi\rangle_D: \nonumber\\
&=&\int d^3p\sum_{\sigma,\sigma'}  k_{i\,\sigma\sigma'}({\bf p})\left[{\frak a}^{\dag}_{\sigma}({\bf p}){\frak a}_{\sigma'}({\bf p}) +{\frak b}^{\dag}_{\sigma}({\bf p}){\frak b}_{\sigma'}({\bf p})\right]\nonumber\\
&&+{i}\int d^3p\, E(p) \sum_{\sigma}\left[{\frak a}^{\dag}_{\sigma}({\bf p}){\tilde\partial}_{i}{\frak a}_{\sigma}({\bf p})+\,{\frak b}^{\dag}_{\sigma}({\bf p}){\tilde\partial}_{i}{\frak b}_{\sigma}({\bf p})\right] \,, \label{Kt0}
\end{eqnarray} 
which depend on the matrices
\begin{eqnarray}\label{Kip}	
 k_{i}({\bf p})&=&\frac{1}{2(E(p)+m)}\epsilon_{ijk}p^j \Sigma_{k}({\bf p})\,.
\end{eqnarray}
The operators (\ref{Kt0}) are self-adjoint but they are not conserved satisfying the  commutation relations of the $sl(2,\mathbb{C})$ algebra,
\begin{equation}
	\left[\mathsf{H},\mathsf{K}_i\right]=-i\mathsf{P}^i\,, \quad  \left[\mathsf{P}^i,\mathsf{K}_j\right]=-i \delta^i_j\mathsf{H}\,,
\end{equation}
 and evolving as 
\begin{equation}
	\mathsf{K}_i(t)=\mathsf{U}(t)\mathsf{K}_i \mathsf{U}^{\dagger}(t) =\mathsf{K}_i+ \mathsf{P}^i\,t\,.
\end{equation} 
On the other hand, we must specify that the operators $\mathsf{K}_i(t)$ cannot be split as the total angular momentum. They  satisfy the canonical relations
\begin{eqnarray}
	\left[\mathsf{J}_i,\mathsf{K}_j(t)\right]&=&i\epsilon_{ijk}\mathsf{K}_k(t)
	\,, \\ 	
	\left[\mathsf{K}_i(t),\mathsf{K}_j(t)\right]&=&-i\epsilon_{ijk}\mathsf{J}_k\,,
\end{eqnarray}  
but without giving relevant commutators with $\mathsf{L}_i$ or $\mathsf{S}_i$. This means that the splitting (\ref{split}) cannot be extended to the entire $sl(2,\mathbb{C})$ algebra.

We derived thus the self-adjoint basis-generators of a family of unitary reps. of the group $\tilde{\cal P}^{\uparrow}_{+}$ with values in a set of one-particle operators which  are determined by the  bases of Pauli spinors $\xi$ we chose for describing polarization. 

\subsection{Coordinate operators}

For performing the quantization of  Pryce's coordinate operator we start with the expectation values derived, according to Eqs. (\ref{ActX}) and (\ref{ActV}),   as
\begin{eqnarray}
\langle \psi, X^i\psi\rangle_D&=&\langle\alpha,\tilde\partial_i\alpha\rangle	+\langle\beta,\tilde\partial_i\beta\rangle \,,\\
\langle \psi, V^i\psi\rangle_D&=&\int d^3p \frac{p^i}{E(p)}\left[\alpha^+({\bf p})\alpha({\bf p}) +\beta^+({\bf p})\beta({\bf p})\right]\,.~~~~\nonumber\\
\end{eqnarray}
As usual, the quantization changes the sign of antiparticle terms such that we obtain the one-particle operators
\begin{eqnarray}
\mathsf{X}^i&=&:\langle\psi, X^i\psi\rangle_D:=\mathsf{X}_+^i+\mathsf{X}_-^i\nonumber\\
&=&i\int d^3p \sum_{\sigma}\left[{\frak a}^{\dag}_{\sigma}({\bf p}) \tilde\partial_i{\frak a}_{\sigma}({\bf p}) - {\frak b}^{\dag}_{\sigma}({\bf p}) \tilde\partial_i{\frak b}_{\sigma}({\bf p})\right]	\,,\label{XX}\\
\mathsf{V}^i&=&:\langle\psi, V^i\psi\rangle_D:=\mathsf{V}_+^i+\mathsf{V}_-^i\nonumber\\
&=&\int d^3p \frac{p^i}{E(p)} \sum_{\sigma}\left[{\frak a}^{\dag}_{\sigma}({\bf p}) {\frak a}_{\sigma}({\bf p}) - {\frak b}^{\dag}_{\sigma}({\bf p}) {\frak b}_{\sigma}({\bf p})\right]\,.\label{VVV}
\end{eqnarray}
We have thus the surprise to see  that the would be mass-center operator proposed by Pryce becomes under quantization the charge-center one or, in other words, the {\em dipole} operator at the time $t=0$. Accordingly, we understand that $\mathsf{V}^i$ are the comps. of the classical current  vector-operator.

This interpretation is confirmed by the commutation relations we briefly inspect in what follows. We start  with 
\begin{equation}
	[\mathsf{H}, \mathsf{X^i}]=-i\mathsf{V}^i\,, \quad 	[\mathsf{H}, \mathsf{V}^i]=0\,,
\end{equation}
showing that the comps. $\mathsf{V}^i$ are conserved while the dipole ones evolve as
\begin{equation}
	\mathsf{X}^i(t)=\mathsf{U}(t)\mathsf{X}^i\mathsf{U}^{\dag}(t)=\mathsf{X}^i+\mathsf{V}^i t\,.
\end{equation}
Moreover, we can verify that $\mathsf{X}^i(t)$  are comps. of a $SO(3)$ vector-operator,
\begin{equation}
\left[\mathsf{L}_i , \mathsf{X}^j(t) \right]=i\epsilon_{ijk} \mathsf{X}^k(t) \,, \quad 	\left[\mathsf{S}_i , \mathsf{X}^j(t) \right]=0\,,
\end{equation} 
which satisfy the canonical relations coordinate-momentum,
\begin{equation}\label{Xcan}
	\left[ 	\mathsf{X}^i(t), \mathsf{X}^j(t) \right]=0\,,\quad	\left[ 	\mathsf{X}^i(t), \mathsf{P}^j \right]=i\delta_{ij}\mathsf{Q}\,,
\end{equation}
in accordance with Eq. (\ref{XPij}) and our interpretation as through quantization $1\in \rho_D$ becomes the charge operator $\mathsf{Q}$. 

However, we did not say which may be the real mass-center operator. Even though there are many definitions of mass-center we can construct easily only  the version of position vectors weighted by rest masses (as in definition (a) of Ref. \cite{B})  defining {\em ad hoc} its comps.  and those of the mass-center velocity as
\begin{equation}\label{XVmc}
	\mathsf{X}^i_{MC}=	\mathsf{X}_+^i-\mathsf{X}_-^i \,,\quad 		\mathsf{V}^i_{MC}=	\mathsf{V}_+^i-\mathsf{V}_-^i\,,
\end{equation}
such that $[\mathsf{H},\mathsf{X}_{MC}^i]=-i \mathsf{V}_{MC}^i$ gives the inertial motion  
\begin{equation}
\mathsf{X}_{MC}^i(t)=\mathsf{U}(t)\mathsf{X}_{MC}^i\mathsf{U}^{\dag}(t)=\mathsf{X}_{MC}^i+\mathsf{V}_{MC}^i t\,.	
\end{equation}
These operators satisfy similar commutation relations as the dipole one apart from the last of  Eqs. (\ref{Xcan}) which reds now
\begin{equation}
	\left[ 	\mathsf{X}_{MC}^i(t), \mathsf{P}^j \right]=i\delta_{ij}\mathsf{N}\,,
\end{equation}
indicating that they describe the kinematics of the center of rest masses which are the same for particles and antiparticles of any momenta. I view of the above results we are skeptical that other coordinate operators satisfying simultaneously similar commutation relations could be derived. Nevertheless, we do not exclude the possibility of finding new  mass-center operators relaxing the canonical conditions as, for example, in the case of spin induced non-commutativity \cite{D}.

The problem which remains open is how the corresponding mass-center operator of RQM may be defined in ${\bf x}$ and ${\bf p}$ reps.. Our preliminary calculations indicate that there exist a spectral rep. solving this problem but the calculations are quite complicated exceeding this paper. We hope to discuss this problem and other versions of mass-center in a further investigation. 

\section{Example: momentum-helicity basis}

The only peculiar polarization used so far is the helicity giving rise to the momentum-helicity basis in which the spinors $\xi_{\sigma}({\bf p})$ and $\eta_{\sigma}({\bf p})=i\sigma_2 \xi^*_{\sigma}({\bf p})$ satisfy the related eigenvalues problems  
\begin{equation}\label{snpp1}
	\hat s_i  {n}^i_{p}\xi_{\sigma}({\bf p})	=\sigma\, \xi_{\sigma}({\bf p})~ \to~ 
	\hat s_i  {n}^i _{p}\eta_{\sigma}({\bf p})	=-\sigma\, \eta_{\sigma}({\bf p}),
\end{equation}
where ${\bf  n}_{p}=\frac{\bf p}{p}$ is the unit vector of ${\bf p}$. One obtains these spinors   transforming the spin basis (\ref{etapm}) as 
\begin{equation}
\xi_{\sigma}({\bf p})=\hat r_h({\bf p}) \xi_{\sigma}~~\to~~\eta_{\sigma}({\bf p})=\hat r_h({\bf p}) \eta_{\sigma}\,,	
\end{equation}
with the help of the $SU(2)$ rotation 
\begin{equation}
	\hat r_h({\bf p})=\sqrt{\frac{p+p^3}{2p}}\left[1_{2\times2}-i \frac{p^1\sigma_2-p^2\sigma_1}{p+p^3}\right]\,.
\end{equation}
The associated $SO(3)$ rotation, $R(\hat r_h({\bf p}))$, transforms the polarization direction ${\bf e}_3$ of the spin basis into the helicity one, ${\bf n}_p$.  In our framework we find that the corresponding  transf. of the mode spinors  is performed by  the integral operator ${\frak R}_h$ whose kernel has the Fourier transform 
\begin{equation}
\tilde {\frak R}_h({\bf p})=	 l_{\bf p}   r_h({\bf p})  l_{\bf p}^{-1}\tilde \Pi_+({\bf p}) + l_{\bf p}^{-1}   r_h({-\bf p})  l_{\bf p}\tilde \Pi_-({\bf p})\,,
\end{equation}
where $ r_h({\bf p})={\rm diag}(\hat  r_h({\bf p}),\hat  r_h({\bf p}))\in \rho_D$.  This trnsf. is complicated but can be controlled by using algebraic codes on computer.

\subsection{Principal operators} 

In this basis the momentum and energy operators which do not depend on polarization lead to the one-particle operators  (\ref{Hom}) and respectively (\ref{Pom}). In contrast, the polarization operator (\ref{Pol}) with  ${\bf n}({\bf p})={\bf n}_{p}$ can be put in the specific form 
\begin{equation}\label{cucu}
\tilde W({\bf p})= s_i  {n}^i_{p}\left[\tilde	\Pi_+({\bf p})-\tilde	\Pi_-({\bf p})  \right]=s_i  {n}^i_{p} \frac{\tilde H_D({\bf p})}{E(p)}\,,
\end{equation}
as ${\bf n}(-{\bf p})=-{\bf n}_p$ and $s_i  {n}^i_{p}$ commutes with $l_{\bf p}$. Then the mode spinors constructed using helicity  spinors satisfy the eigenvalues problems (\ref{WU}) and (\ref{WV}) which guarantee the convenient quantization  (\ref{Polq}). 

On the other hand, here we may use the components of the Pauli-Lubanski operator interpreted as a covariant four-vector spin operator.  Its 0-th component is the helicity operator 
\begin{equation}\label{mucu}
W_0= J_i P^i=s^i P^i ~\Rightarrow~  \tilde W_0({\bf p})=\tilde S_i ({\bf p})p^i=s_i p^i\,,
\end{equation} 
whose action on the mode spinors,
\begin{eqnarray}
(W_0U_{{\bf p}, \xi_{\sigma}({\bf p})})(x)&	=&\tilde W_0({\bf p})U_{{\bf p}, \xi_{\sigma}({\bf p})}(x)=\sigma p U_{{\bf p}, \xi_{\sigma}({\bf p})}(x)\,,\\
(W_0V_{{\bf p}, \eta_{\sigma}({\bf p})})(x)&	=&\tilde W_0(-{\bf p})V_{{\bf p}, \eta_{\sigma}({\bf p})}(x)=\sigma p V_{{\bf p}, \eta_{\sigma}({\bf p})}(x)\,,\label{Vhel}
\end{eqnarray}
calculated as in the previous case,  is different from that of $W$ as the eigenvalue of Eq. (\ref{Vhel}) is $\sigma p$ instead of $-\sigma p$.  This difference  comes from the term $E(p)^{-1}\tilde  H_D({\bf p})$ of Eq. (\ref{cucu})  which assures suitable eigenvalues of the operator $W$. However, the inverse antiparticle eigenvalues of the helicity operator are not a major impediment such that in QFT we may  use either the operator $ \mathsf{W}$ defined by Eq. (\ref{Polq}) or the helicity one,
\begin{equation}\label{W0}
\mathsf{W}_0=\int d^3p\,p\sum_{\sigma} \sigma\left[{\frak a}^{\dag}_{\sigma}({\bf p}){\frak a}_{\sigma}({\bf p}) -{\frak b}^{\dag}_{\sigma}({\bf p}){\frak b}_{\sigma}({\bf p})\right]	\,.,
\end{equation}
bearing in mind its specific action.
 
For writing down the spin comps. (\ref{Spin})  we derive the matrices (\ref{Dxx})  in this basis, 
\begin{eqnarray} 
\Sigma_{1}({\bf p})&=&\frac{p^1}{p}\,\sigma_3 -p^1\frac{p^1 \sigma_1+p^2\sigma_2}{p(p+p^3)}+\sigma_1\,,\nonumber\\
\Sigma_{2}({\bf p})&=&\frac{p^2}{p}\,\sigma_3 -p^2\frac{p^1 \sigma_1+p^2\sigma_2}{p(p+p^3)}+\sigma_2\,,\label{Sigp}\\
\Sigma_{3}({\bf p})	&=&\frac{p^3}{p}\sigma_3-\frac{p^1 \sigma_1+p^2\sigma_2}{p}\,,\nonumber
\end{eqnarray}
which satisfy $p^i\Sigma_i({\bf p})=p\sigma_3$. The form of the 'covariant' derivatives $\tilde\partial_i=\partial_{p^i} 1_{2\times2}+\Omega_i({\bf p})$ is determined by  the matrices (\ref{Omega}) that read
\begin{eqnarray}
\Omega_1({\bf p})&=&\frac{-i}{2p^2(p+p^3)}\left[ p^1p^2\sigma_1 +pp^2\sigma_3\right.\nonumber\\
&&\left.\hspace*{12mm}+(pp^3+{p^2}^2+{p^3}^2)\sigma_2\right]\,, \nonumber\\
\Omega_2({\bf p})&=&\frac{i}{2p^2(p+p^3)}\left[ p^1p^2\sigma_2 +pp^1\sigma_3\right.\label{Omegp}\\
&&\left.\hspace{12mm}+(pp^3+{p^1}^2+{p^3}^2)\sigma_1\right]\,,\nonumber\\
\Omega_3({\bf p})&=&\frac{i}{2p^2}\left(	p^1\sigma_2-p^2\sigma_1\right)\,.\nonumber
\end{eqnarray}
and satisfy $p^i\Omega_i({\bf p})=0$. We obtain thus  apparently complicated  matrices $\Sigma_i$ and $\Omega_i$ but whose algebra is the same as in momentum-spin basis where $\Omega_i=0$ and $\Sigma_i=\sigma_i$. The identities  (\ref{idimidi}) help us to show that  $\Sigma_j$ and $\tilde\partial_{i}$ satisfy the same commutation relations as $\sigma_i$ and  $\partial_{p^i}$. Using  these matrices we may derive for the first time all the isometry generators and kinetic operators acting on the Fock state space of QFT in momentum-helicity basis. 

\subsection{One-particle relativistic quantum mechanics}

In applications we may turn back to RQM but considered now as the one-particle restriction of QFT. For example, in the normalized one-particle state 
\begin{equation}
|\alpha\rangle=\int d^3 p \sum_{\sigma} \alpha_{\sigma}({\bf  p}){\frak a}^{\dagger}_{\sigma}({\bf p})|0\rangle \,, \quad \langle \alpha|\alpha \rangle=1\,,
\end{equation}
defined by the wave functions $\alpha_{\sigma}({\bf p} )$ that form the normalized Pauli spinor $\alpha\in{\cal F}_{\alpha}$ as in Eq. (\ref{alpha}), we may calculate the expectation value of any generator $\mathsf{X}$ as
\begin{equation}
\langle \alpha|\mathsf{X}|\alpha\rangle =\langle \alpha , \tilde X \alpha\rangle\,,	
\end{equation} 
where $\tilde X\in {\rm Lie}(\tilde T)$ is the generator of RQM  corresponding to $\mathsf{X}$. The list of these generators can be written down, according to Eqs.  (\ref{Hom}-\ref{Kip}), but omitting the explicit dependence on ${\bf p}$,  
\begin{eqnarray}
\tilde P^i&=&p^i\,, \qquad~~~ \tilde H=E\,,\quad~~~~ \tilde W=\frac{1}{2}\sigma_3\,,~~~~\tilde W_0=\frac{p}{2}\sigma_3\,,\nonumber\\
\tilde J_i&=&\tilde L_i+\tilde S_i\,,\quad\tilde S_i= \frac{1}{2}\,\Sigma_i\,, \quad \tilde L_i=-i\epsilon_{ijk}p^j\tilde\partial_{k}	\,,\label{tab}\\
\tilde K_i&=& i E\tilde\partial_{i}+\frac{1}{2(E+m)}\,\epsilon_{ijk}p^j \Sigma_{k}\,.\nonumber
\end{eqnarray}  
Similarly we find $\tilde Q=1$ and the kinetic operators
\begin{equation}
\tilde{X}^i=\tilde{X}^i_{MC}=i \tilde{\partial}_i\,,\quad	\tilde{V}^i=\tilde{V}^i_{MC}=\frac{p^i}{E}\,.
\end{equation}
of the particle inertial motion.

More algebra may be performed  resorting to algebraic codes on computer. We obtain thus the space comps. of the Pauli-Lubanski operator, 
\begin{equation}
	\tilde W_i=E \tilde J_i+\epsilon_{ijk}p^j \tilde K_k=m\tilde S_i+\frac{p^i}{E+m} \tilde W_0\,, 
\end{equation}
and verify that the second Casimir operator (\ref{CM2}) in this rep., gives the expected invariant $\tilde C_2=-\eta^{\mu\nu}\tilde W_{\mu}\tilde W_{\nu}=\frac{3}{4} m^2$.  Note that in this rep. the first  invariant (\ref{CM1}) is implicit as $E$ is just the relativistic energy. 

This example shows that the structure and properties of the spin, angular momenta and polarization operators defined here are close to  the  original non-relativistic Pauli's spin theory in ${\bf p}$-rep.. However, this happens only in the particle sector as for antiparticles there is a discrepancy because  of  the operators $\tilde Q, \tilde W_0$, $\tilde X^i $ and $\tilde V^i$ which change their signs. 

\section{Concluding remarks}

We have shown that the comps. of spin operator proposed by Pryce in ${\bf p}$-rep. are  Fourier transforms of the kernels of  integral operators generating a rep. of the little group $SU(2)$ carried by the space of Pauli's spinors determining the polarization. Therefore, these operators are conserved via Noether theorem, allowing us to define the conserved polarization operator (\ref{Pol}).

After quantization  the spin operator becomes the desired conserved one-particle operator of comps. (\ref{Spin}),  splitting naturally the total angular momentum  in two conserved parts, i. e.  this spin operator and the associated conserved angular momentum of comps. (\ref{Lang}). Therefore, we must accept that this is the correct spin operator we need for defining and controlling the polarization in special relativistic QFT. The action of the corresponding polarization operator  (\ref{Polq}) defined here for the first time confirms this interpretation.

In contrast, the associated Pryce coordinate operator defined initially in ${\bf p}$-rep. as a mass-center one becomes after quantization the dipole operator (\ref{XX}) evolving linearly in time tanks to  the conserved classical current (\ref{VVV}). Nevertheless, a mass-center position and velocity operators (\ref{XVmc}) can be written by hand but corresponding to   another definition of mass-center, different from that considered by Pryce \cite{B}.   Thus we verify again that the correct physical meaning of the relativistic observables can be found only in QFT. 

As an application we derived the matrices (\ref{Sigp}) and (\ref{Omegp}) we need for writing down the isometry generators and kinetic operators in momentum-helicity basis.  Turning back to the RQM seen as a one-paricle restriction of QFT we obtained the operators (\ref{tab}) that are close to those of the original non-relativistic Pauli theory. We believe that this is a significant physical argument in favor of our approach. 

Unfortunately, we do not have other examples of peculiar polarization as the helicity is the only one used so far. We hope that our  spin and polarization operators defined here will offer one the opportunity of defining new types of peculiar polarization that could be observed in further experiments.

\subsection*{Appendix A:  Boosts and Foldy-Wouthuysen transformations}

\setcounter{equation}{0} \renewcommand{\theequation}
{A.\arabic{equation}}

The standard Lorentz boosts of  $\rho_D$ are transfs. of the form (\ref{l}) with parameters $\tau^i=-\frac{p^i}{p}{\rm tanh}^{-1} \frac{p}{E(p)}$ that read  \cite{Th}
\begin{eqnarray}\label{Ap}
	l_{{\bf p}}=\frac{E(p)+m+\gamma^0\gamma^i p^i}{\sqrt{2m(E(p)+m)}}\in \rho_D\,,
\end{eqnarray}
giving rise to the boosts $L_{\bf p}=\Lambda(l_{\bf p})\in L_{+}^{\uparrow}$ with the matrix elements \cite{WKT}
\begin{eqnarray}
	(L_{{\bf p}})^{0\,\cdot}_{\cdot\, 0}&=&\frac{E(p)}{m}\,,\quad (L_{{\bf p}})^{0\,\cdot}_{\cdot\, i}=(L_{{\bf p}})^{i\,\cdot}_{\cdot\, 0}=\frac{p^i}{m}\,,\nonumber\\ 
	(L_{{\bf p}})^{i\,\cdot}_{\cdot\, j}&=&\delta_{ij}+\frac{p^i p^j}{m(E(p)+m)}\,.\label{Lboost}
\end{eqnarray}
The matrices (\ref{Ap})satisfy $l_{\bf p}=l_{\bf p}^+$ and $l^{-1}_{\bf p}=l_{-\bf p}=\gamma^0l_{\bf p}\gamma^0$ and  
\begin{equation}\label{Ap2}
l_{\bf p}^2=\frac{E({p})+\gamma^0\gamma^i p^i}{m}\,,\qquad 	l_{-\bf p}^2=\frac{E(p)-\gamma^0\gamma^i p^i}{m}\,,
\end{equation}
giving rise to the following identities
\begin{eqnarray}
	\frac{1+\gamma^0}{2}l_{\bf p}^2	\frac{1+\gamma^0}{2}&=&\frac{E(p)}{m}	\frac{1+\gamma^0}{2}\,,\label{idll}\\ 	
	\frac{1-\gamma^0}{2}l_{-\bf p}^2	\frac{1-\gamma^0}{2}&=&\frac{E(p)}{m}	\frac{1-\gamma^0}{2}\,.\label{idlll}
\end{eqnarray}
which help us to recover the integral operators $\Pi_+$ and $\Pi_-$ defined by Pryce  \cite{B} whose kernels have the Fourier transforms.
\begin{eqnarray}
\tilde	\Pi_+({\bf p})&=&\frac{m}{E({p})}l_{\bf p}	\frac{1+\gamma^0}{2} l_{\bf p}=\frac{1}{2}\left(1+\frac{\tilde H_D({\bf p})}{E(p)}\right)\,,\label{Pip}\\ 
\tilde\Pi_-({\bf p})&=&\frac{m}{E({p})}l^{-1}_{\bf p}	\frac{1-\gamma^0}{2} l^{-1}_{\bf p}=\frac{1}{2}\left(1-\frac{\tilde H_D({\bf p})}{E(p)}\right)\,,\label{Pim}
\end{eqnarray}
where $\tilde H_D({\bf p})$ is given by Eq. (\ref{HDp}). It is not difficult to verify that $\Pi_+$ and $\Pi_-$ form a complete system of  orthogonal projection operators satisfying $\Pi_+^2=\Pi_+\,,~ \Pi_-^2=\Pi_-\,,~ \Pi_+\Pi_-=\Pi_-\Pi_+=0$ and $\Pi_+ +\Pi_-=1\in \rho_D$. According to Eqs. (\ref{HDU}) and (\ref{HDV}) we find that these operators separate the mode spinors of positive and negative frequencies as $\Pi_+{\cal F}_D={\cal F}_D^+$ and  $\Pi_-{\cal F}_D={\cal F}_D^-$ \cite{B}. 

Looking for a unitary transf. able to bring the  Hamiltonian $\tilde H_D({\bf p})$ in diagonal form, Foldy and Wouthuysen found the unitary transf. \cite{FW}
\begin{equation}\label{FW}
	U_{FW}({\bf p})=U_{FW}^+(-{\bf p})=\frac{E(p)+m+\gamma^i p^i}{\sqrt{2E(p)(E(p)+m)}}  
\end{equation}
acting as
\begin{equation}\label{FW1}
U_{FW}({\bf p})	\tilde H_D({\bf p})U_{FW}(-{\bf p})=\gamma^0 E(p)\,.
\end{equation}
Assuming that in the frame where the Hamiltonian is diagonal the spin operator is the Pauli one, ${\bf s}$, and applying the inverse transf.,  
\begin{equation}\label{FW3}
\tilde {\bf S}({\bf p})=U_{FW}(-{\bf p}){\bf s}\,U_{FW}({\bf p})\,,
\end{equation}
it turns out just the Pryce spin operator whose comps.  are given by Eq. (\ref{Sip}). For this reason $\tilde {\bf S}({\bf p})$ is called often the Foldy-Wouthuysen spin operator. In fact the spin operator is the same but defined in two different manners: either indirectly in association with Pryce's coordinate operator or through the transf. (\ref{FW3}). Note that both these definitions are different from that we propose here for the same operator.

\subsection*{Appendix B:  Induced representatios.}

\setcounter{equation}{0} \renewcommand{\theequation}
{B.\arabic{equation}}

The induced reps. are a tool for constructing unitary reps. of a local-compact group in terms of unitary ones of a compact subgroup \cite{Wig,Mc}. Given a local-compact group ${G}$,  a subgroup ${H}$ and the function $\phi : {G} \to {\cal V}$, with values in a vector space ${\cal V}$, one says that the natural rep. $\pi(g)\phi(x)=\phi(g^{-1}x)$ is induced by the rep. $\tau$ of the group $H$ if \cite{Th}
\begin{equation}\label{indu}
	\phi(x h^{-1})=\tau(h)\phi(x) \,,\quad   \forall\, x\in G\,,~~~ h\in H\,.
\end{equation}
Bearing in mind that a Haar measure can be defined at any time on the cosset space $G/H$  it is convenient to consider the new function $\hat\phi=\phi\circ\chi: G/H\to{\cal V}$ defined with the help of an arbitrary function $\chi: G/H\to G$. If  $\tau$ is an unitary rep. then the induced rep. is unitary transforming the functions $\hat\phi  \in {\cal L}^{2}({G}/{H},{\mu},{\cal V})$ but preserving the scalar product of this Hilbert space \cite{Mc}.  An induced rep. is irreducible if the rep. $\tau$ is irreducible. When fermions must be studied then  instead of $G$ and $H$ we consider their universal covering groups,  $\tilde G$ and the little  group $\tilde H$ \cite{Th}.

In RQM the wave functions of ${\bf p}$-rep. transform under translations by simple multiplications with phase factors such that we may restrict ourselves to the reps. of the groups $G=L_{+}^{\uparrow}$ or $\tilde G=SL(2,\mathbb{C}) $. The wave functions are defined on orbits associated to  representative momenta as in Sec. 3.2. Each orbit $\Omega_{\mathring p}$ is isomorphic with the cosset space $L_{+}^{\uparrow}/H$ where $H$ is the stable group of the representative momentum $\mathring p$. Therefore,  we can define the mapping $\chi: \Omega_{\mathring p} \to G$ such that $p=\chi(p)\mathring p$. Then the natural action of  any $\Lambda\in   L_{+}^{\uparrow}$ can be written as 
\begin{equation}
\Lambda: \hat \phi(p)\to \phi\left(\Lambda^{-1} \chi(p)\right)=\phi\left(\chi(\Lambda^{-1}p) W(\Lambda, p)^{-1} \right)\,,
\end{equation}
where $W(\Lambda,p)=\chi^{-1}(p)\Lambda\chi(\Lambda^{-1} p)\in H$ is a Wigner transf. of  stable group,  as $W(\Lambda,p) \mathring p =\mathring p$. Finally, we find the transformation rule 
\begin{equation}\label{ind1}
\Lambda: \hat \phi(p)\to\tau(W(\Lambda,p))\hat\phi(\Lambda^{-1} p)	
\end{equation}
resulted from condition (\ref{indu}).

For massive fermions discussed in Sec. 3.2 we chose $\chi(p)=L_{\bf p}$  defined by Eqs. (\ref{Lboost}). In this case the little group $\tilde H=SU(2)$  is compact allowing finite-dimensional unitary and irreducible reps.. We use here only the rep. of spin $\frac{1}{2}$ such that  $\sqrt{E}\alpha$ and $\sqrt{E}\beta$ are functions of the Hilbert space ${\cal L}^2(\Omega_{\mathring p}, \mu,{\cal V}_P)$ with the invariant measure  (\ref{measure}). These  functions  transform as in Eq. (\ref{ind1}) such that the functions $\alpha$ and $\beta$ transform according to the rule (\ref{Wig}) deduced indirectly from Wigner's approach.  Thus for massive particles the covariance is solved in terms of unitary reps. with a natural physical meaning.

However, the group $SO(3)$ is the only compact subgroup of the group $L_{+}^{\uparrow}$ such that for other orbits there are difficulties. In the case of massless particles  the representative momentum $\mathring p=(1,0,0,1)$ has the stable group $H=E(2)$ formed by $SO(2)$ rotations and two nilpotent translations \cite{nul} whose effect must be eliminated in order to keep only  the unitary reps. of the subgroup $SO(2)$. This can be done resorting to some supplemental  restrictions, e. g. keeping only the left-handed components of neutrino or setting the Coulomb gauge of  Maxwell field. For tachyons having  the stable group $H=SO(2,1)$ there are no solutions.

\end{document}